\documentclass[%
 reprint,
superscriptaddress,
 amsmath,amssymb,
 aps,
 tabularx,
 siunitx,
]{revtex4-2}
\usepackage{graphicx}
\usepackage{dcolumn}
\usepackage{bm}
\usepackage[dvipsnames]{xcolor}
\usepackage{siunitx}
\usepackage[version=4]{mhchem}

\newcommand{\change}[1]{\textcolor{black}{#1}}

\begin{document}

\preprint{APS/123-QED}

\title{Population effects driving active material degradation in intercalation electrodes}

\author{Debbie Zhuang}
\email{dezhuang@mit.edu}
\affiliation{Department of Chemical Engineering, Massachusetts Institute of Technology \\ 77 Massachusetts Avenue, Cambridge, MA 02139}
\author{Martin Z. Bazant}
\email{bazant@mit.edu}
\affiliation{Department of Chemical Engineering, Massachusetts Institute of Technology \\ 77 Massachusetts Avenue, Cambridge, MA 02139}
\affiliation{Department of Mathematics, Massachusetts Institute of Technology \\ 77 Massachusetts Avenue, Cambridge, MA 02139}

\date{\today}

\begin{abstract}
\change{In battery modeling, the electrode is discretized at the macroscopic scale with a single representative particle in each volume.}
This lacks the accurate physics to describe interparticle interactions in electrodes.
To remedy this, we formulate a model that describes the evolution of degradation of a population of battery active material particles using ideas in population genetics of fitness evolution, where the state of a system depends on the health of each particle that contributes to the system.
With the fitness formulation, the model incorporates effects of particle size and heterogeneous degradation effects which accumulate in the particles as the battery is cycled, accounting for different active material degradation mechanisms.
At the particle scale, degradation progresses nonuniformly across the population of active particles, observed from the autocatalytic relationship between fitness and degradation. Electrode-level degradation is formed from various contributions of the particle level degradation, especially from smaller particles.
It is shown that specific mechanisms of particle level degradation can be associated with characteristic signatures in the capacity-loss and voltage profiles.
Conversely, certain features in the electrode-level phenomena can also provide insight into the relative importance of different particle level degradation mechanisms.

\end{abstract}

\maketitle

\section{Introduction}


Lithium ion batteries are driving a revolution in electronic devices and electric vehicles for a more sustainable, electrified future \cite{li201830}.
Degradation in lithium-ion batteries with aging is a major problem that prevents global electrification by shortening the lifetime of Li-ion batteries \cite{kabir2017degradation, goodenough2013li, berliner2021nonlinear}.
Cathode degradation is generally caused by particle level physical mechanisms such as electrochemical resistance growth from films at the electrode/electrolyte interface \cite{attia2019electrochemical, das2019electrochemical}, phase transformations and loss of kinetic abilities at the surface or bulk \cite{yan2015evolution, zheng2019ni}, as well as \change{electrolyte loss} \cite{sarasketa2015understanding, nanda2020lithium}.
\change{The apparent capacity loss from the convolution of these mechanisms cannot be physically explained by a single degradation mechanism, since it is caused by many degradation mechanisms, and results in slightly different behavior in electrode-level phenomena.}

Electrode-level degradation phenomena is seen in capacity loss curves and in voltage-capacity curves \cite{severson2019data}.
This is seen through impedance experiments, where impedance growth has been found to be mainly from the cathode side \cite{ma2019hindering, attia2022knees, nagasubramanian2000impedance, sallis2016surface}, while anode degradation is generally motivated by solid electrolyte interphase formation and lithium plating \cite{gao2021interplay, attia2019electrochemical}.
Understanding the increase of cathode impedance is a critical step of deconvoluting electrode-level degradation.
Previous work \cite{ma2019hindering, mandli2019analysis} has identified resistance growth as a large component of this failure, but has not clarified the separation of all the different mechanisms.
To deconvolute these particle level mechanisms, it is imperative to understand the different relationships between the particle level driving forces and electrode-level behavior of the degradation mechanisms, especially in relation to particle population dynamics.
This reveals a more complete understanding of electrode degradation.

In many physical theories, single particle models are used, but can be inaccurate due to not accounting for interactions between particles.
This concerted behavior between individuals in population dynamics has been observed in many systems, from biological systems such as fireflies \cite{ott2017frequency, kuramoto1975international} to electrochemical oscillations in batteries \cite{sun2018electrochemical}.
In a battery, the effects of population dynamics can \change{appear in solid solution materials \cite{park2021fictitious} as well as in phase separating materials as each particle activates and phase transitions} \cite{ferguson2012nonequilibrium, ferguson2014phase, li2014current, lim2016origin}.
Past biological modeling used population dynamics \cite{bacaer2011short, herrmann2012kramers, herrmann2014rate} studied with the Fokker-Planck equation \cite{kolmogorov1study} to understand the growth and eventual death of biological populations \cite{fisher1958genetical, wright1984evolution}.
The idea of a fitness landscape \cite{wright1984evolution}, where the fitness represents the reproductive rate of a genotype, or ``effectiveness'', was incorporated into these models to explain why populations evolve towards certain traits.
We can similarly apply this idea to model particle population dynamics in lithium-ion batteries.
\begin{figure*}[t]
\includegraphics[width=\textwidth]{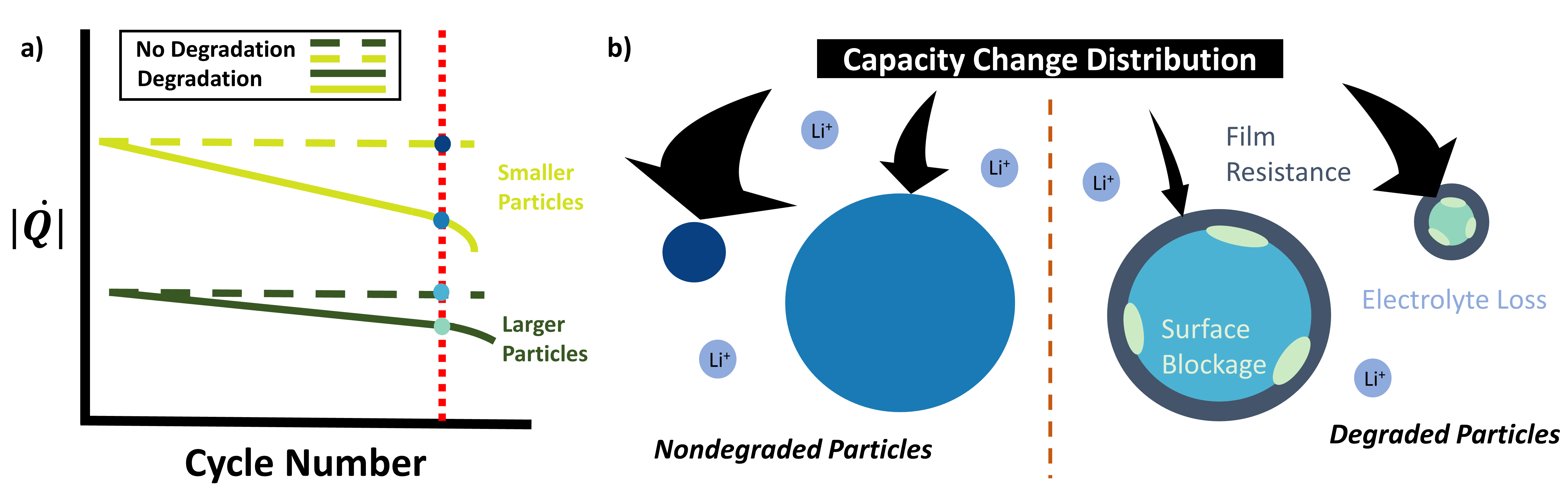}
\centering
\caption{Schematic of particle size distribution effect on current distribution in a constant current charge or discharge simulation for a battery.
a) The absolute value of the capacity change fraction of each particle is plotted with respect to the cycle number as the system is degraded.
This accounts for effects of degraded charge transfer kinetics and particle size contributions by reducing the nondegraded current to the degraded current.
\change{b) A snapshot at a single time point for a sample of degraded or nondegraded particles of small or large particle sizes is shown, where the capacity change distribution splits the current between the different particles based on their size effects and degradation.}}
\label{fig:fig0}
\end{figure*}

For a battery, the fitness can be envisioned for each particle as the effectiveness of carrying the current load, as shown in Fig. \ref{fig:fig0} for a population distribution affected by degradation effects and particle size effects.
Fig. \ref{fig:fig0}a shows the fitness as a rate of change of the capacity fraction of each particle, which is the capability of carrying current.
However, fitness landscapes can also evolve depending on the environment they are in \cite{fisher1958genetical, desai2007beneficial}, which in turn affects the evolution of population dynamics, called coevolution \cite{mustonen2009fitness}.
In a system of battery particles, the fitness landscape of the particles changes \cite{guha2017remaining} as the battery is cycled, similarly to how the fitness of different genotypes changes with evolution \cite{kauffman1991coevolution}.
Thus, it is important to quantify  the fitness landscape of battery particle systems as they evolve.
In this work, through methods of population dynamics coupled with electrochemical kinetics in a simple battery model, we gain an understanding of active material degradation at the particle scale and how it affects electrode-level degradation phenomena.
We simulate an interacting population of particles in a single electrode volume in a porous electrode.
Initially, particle size effects influence the current distribution in the population,
while as the battery is cycled, degradation also influences the current distribution.
Fitness and degradation are observed to have an autocatalytic relationship \change{from the simulations}.
We use specific degradation mechanisms to understand the effect on the electrode-level behavior of each of these materials.
Inversely, experimental information on the capacity loss profile and voltage-capacity curves may provide insight on the particle level degradation mechanisms present.
%

\section{Theory}

In porous electrode theories, single particle models are often used to described the active material interactions of the system \cite{newman2012electrochemical}, which can be inaccurate from not accounting for interparticle interactions.
To remedy this, we formulate a theory
in porous intercalation electrodes 
featuring the cumulative effect of population dynamics including degradation.
We consider particles with a size distribution, each of which is treated homogeneously during (de)intercalation. 
The population $f(t,c;r)$ is the probability distribution of particles with radius $r$ and concentration $c$ at time $t$. 
The evolution of population $f(t,c;r)$
can be tracked by the modified Fokker-Planck equation. 
During evolution, the fitness function $W$ is introduced 
to modify the intercalation rate of degradation.
The analytic expressions of fitness functions for three degradation modes are derived. 


\subsection{Conservation Equation}

A porous electrode is normally modeled using volume discretizations at the electrode scale, with single particle models in each volume to simulate the particle level intercalation and diffusion mechanisms of lithium ions \cite{newman2012electrochemical, doyle1993modeling}. 
To account for interparticle interactions, many active material particles need to be modeled in the same volume, which are under the same voltage.
The dynamics of a population can be driven by controlled parameters imposed on the system \cite{ramkrishna2000population, ramkrishna2014population}, such as the applied current or voltage.
Neglecting transport limitations at the electrode scale, these models can be simplified to a single volume.

We can use the Fokker-Planck equation, which is commonly used to simulate population dynamics of evolving probability distributions $f(t,c;r)$ of multiple fields (such as concentration and particle size) under time evolution, to model this as
\begin{equation}
\label{eq:full_FP}
    \frac{\partial f}{\partial t} = - \frac{\partial}{\partial c}\left(fR\right) + \frac{\partial}{\partial c}\left(D \frac{\partial}{\partial c}\left(fW\right)\right).
\end{equation}
In this equation, the first term is from the mean of the reaction happening amongst the particles in the system, while the second term is from the variance of the reaction amongst the particles.
The variables are defined as $R$, the volumetric reaction rate derived from the mean of the transition rates, and $D$, the thermal diffusivity related to the fluctuations of the transition rates \cite{frank2005nonlinear}, \change{with the full definitions in Appendix \ref{ref:appdx_derivation_FP}}.
The full derivation of the modified Fokker-Planck equation accounting for the fitness value can be seen in Appendix \ref{ref:appdx_derivation_FP} and reveals that the appearance of the fitness function $W$ is from the change in the effective reaction rate from degraded charge transfer kinetics and particle size effects \change{\cite{zhao2019population, herrmann2014rate, herrmann2012kramers, dreyer2011behavior}}. 
For better understanding of interparticle effects, we write this in the form where the volumetric reaction is replaced by the nondegraded intercalation rate per area multiplied by the fitness of the battery $W$ as
\begin{equation}
\label{eq:approx_FP}
    \frac{\partial f}{\partial t} = - \frac{\partial}{\partial c}\left(f\bar{i}W\right) + \frac{\partial}{\partial c}\left(D \frac{\partial}{\partial c}\left(fW\right)\right),
\end{equation}
where $D = k k_B T/N_t$ is the thermal diffusivity parameter, $k = \bar{i}/\eta$ for a linear form of the reaction rate, and $N_t$ is the maximum number of molecules in the system. 
Here $\bar{i}$ is the nondegraded intercalation current per area, $i$ is the real intercalation current per area, $R$ is the real volumetric reaction rate per volume, and $W = R/\bar{i}$ is the fitness function, which is defined and expounded in the following section.

\subsection{Intercalation Kinetics}\label{sec:interc_kinetics}

The intercalation kinetics control the mean of the time evolution of the system as defined in Eq. \ref{eq:approx_FP}, and thus an accurate description of kinetics becomes imperative \cite{zhao2019population, yang2019quantification, xu2019heterogeneous}.
For electrochemical systems, well-known physical models exist for thermodynamically consistent reactions such as the Butler-Volmer reaction  \cite{faulkner2002electrochemical, newman2012electrochemical, doyle1993modeling}.
Increasingly complex models as coupled ion electron transfer (CIET) theory \cite{fraggedakis2021theory, Zhang2022_CIET_preprint} have been further derived and verified experimentally from x-ray imaging of lithium iron phosphate materials to account for electron availability from the density of states of the intercalation material.
These intercalation models can be used to quantify the heterogeneity and degradation growth in a population of battery particles.

The general form of the intercalation reaction takes the form of 
\begin{equation}
    i(\eta) = k_0(c) h(\eta); ~~h(\eta) = r_{\leftarrow} - r_{\rightarrow},
\end{equation}
where $k_0(c)$ is a transition state term describing the overall rate of reaction (incorporating exchange current density) and $h(\eta)$ is from the thermodynamic driving force composed of a forward $r_{\rightarrow}$ and backward reaction $r_{\leftarrow}$ driven by the overpotential
\begin{equation}
e\eta = \left( e\phi_s + \mu(c)\right) - \left( e\phi_+ + k_B T \ln{a_+}\right),
\end{equation}
where $c$ is the concentration of the intercalated lithium, $\mu(c)$ is the chemical potential of the intercalated lithium, $\phi_{+/s}$ is the lithium ion electrical potential in electrolyte or solid, and $a_+$ is the activity of the lithium ions in electrolyte.

When there is no degradation accounted for in the system, the intercalation reaction can be similarly defined as
\begin{equation}
    \bar{i}(\eta) = \bar{k}_0(c) \bar{h}(\eta),
\end{equation}
where the overbar indicates a system without degradation.
The Butler-Volmer reaction rate is modeled with transition state theory as
\begin{equation}
\bar{i}(\eta) = k_0 (c)\left( \exp{\left(-\alpha \eta\right)} - \exp{\left((1-\alpha)\eta\right)}\right),
\end{equation}
with the thermodynamically consistent prefactor $k_0(c) = k_0^* c^\alpha (1-c)^{1-\alpha}a_+^{1-\alpha}$, where $k_0^*$ is the reaction current prefactor, $\alpha$ is the charge transfer coefficient, and $a_+$ is the activity coefficient for the electrolyte.
The reaction rate from coupled ion electron transfer, which accounts for electron availability, is modeled using an approximation as \cite{fraggedakis2021theory, zeng2014simple} \begin{equation}
\bar{i}(\eta) = k_0 (c)\left(a_+ \text{helper}(-\eta_f  , \lambda) - c \text{helper}(\eta_f  , \lambda)\right),
\end{equation}
with prefactor $k_0(c) = k_0^* (1-c)$, where the helper function is defined as \cite{zeng2014simple}
\begin{widetext}
\begin{equation}
    \text{helper}(\eta_f, \lambda) = \frac{\sqrt{\lambda \pi}}{1+\exp{(-\eta_f)}} \text{erfc}\left(\frac{\lambda - \sqrt{1+\sqrt{\lambda} + \eta_f^2}}{2\sqrt{\lambda}} \right).
\end{equation}
\end{widetext}
Here, the formal overpotential is defined as $e\eta_f = e\eta - k_B T \ln{\frac{c}{a_+}}$, which is the overpotential with the ionic concentration dependencies removed, and $\lambda$ is the Marcus reorganization energy for electron transfer in the solid material.
\change{In this electrochemical reaction, the overall reaction is driven by the difference between the reduction and oxidation reactions, $\bar{i} = \bar{i}_{red} - \bar{i}_{ox}$, where $\bar{i}_{red} = k_0(c)a_+ \text{helper}(-\eta_f, \lambda)$ and $\bar{i}_{ox} = k_0(c)c \text{helper}(\eta_f, \lambda)$.}
There is a limiting current reached with this model from the density of states of the material used with respect to overpotential, $i_{lim} = k_0^*(1-c)$, unlike the Butler-Volmer model, which grows exponentially with overpotential \cite{Zhang2022_CIET_preprint}.
Kinetic behavior at high overpotentials is strongly affected by the correct choice of reaction kinetics.

An important term we will encounter is the idea of an inverse differential resistance, or \change{differential conductance}, $\frac{\partial \bar{i}}{\partial \eta}$, which is first mentioned in Refs. \cite{bazant2017thermodynamic, bazant2013theory} to explain the idea of autocatalytic reactions. 
Since degradation affects the overpotential,
\change{the differential conductance reveals the acceleration of the reaction with overpotential.}
Experimentally, this \change{differential conductance} also appears in the charge transfer resistance term in electrochemical impedance spectroscopy measurements \cite{faulkner2002electrochemical}.
A negative \change{differential conductance} means that the reaction is self-driving or autocatalytic, as opposed to a reaction that is self-limiting or autoinhibitory.
This \change{differential conductance} incorporates an effect from the driving force term $h(\eta)$ as well as from the transition state term $k_0(c)$.
For a solid solution material, the effect of the driving force term is generally autocatalytic, with the exception of Marcus-type electron transfer reactions, which may have inverted regions causing local values of autoinhibitory reactions \cite{miller1984intramolecular, chidsey1991free}.
However, the effect of the transition state term is not necessarily autocatalytic and can amplify nonheterogeneity in population dynamics, as explored in Ref. \cite{zhao2019population}.

\subsection{Fitness Function}
As seen from the population evolution equation (Eq. \ref{eq:approx_FP}), the fitness of particle population dynamics, especially in relation to resistance evolution as the battery is cycled, is important to quantify \cite{pinson2012theory, nie2013role, das2019electrochemical, attia2019electrochemical}.
The coupling of degradation evolution, kinetics, and particle size distributions in a porous electrode model is difficult to model because capturing multiscale effects \cite{darling1997modeling} and calculating implicit solutions numerically at each step in the Fokker-Planck solution is computationally expensive.
We aim to formulate in a simple manner the solution of the Fokker-Planck equation with the formulation of the fitness function $W$.

The fitness model captures a ratio between the real and ideal currents, which linearizes the kinetics of the system.
The advantage of using a fitness function formulation is that from the definition of the fitness variable, we keep the original formulations of our reaction rates without having to self consistently solve implicit kinetic equations in the population balance.
This becomes especially difficult for models with a film resistance contribution, as implicit solutions are necessary for these models.
In addition, with the use of a fitness function model, there is a separation between the different contributions of particle size as well as modes of degradation on fitness, so there is a clear dominant mechanism from degradation.

The nondegraded intercalation current without degradation $\bar{i}$ is the reference that the current at a certain overpotential reaches with no degradation.
In physical particles, there are often degradation effects, especially as the particles are cycled, so the real current magnitude $|i|$ is generally smaller than the nondegraded current magnitude $|\bar{i}|$. 
The general definition of the fitness, or ``performance'', is the ratio of the volumetric reaction rate accounting for particle size and degradation effects, or
\begin{align}
    W = \frac{R}{\bar{i}},
\end{align}
where $R$ is the volumetric reaction rate and $\bar{i}$ is the nondegraded intercalation current per area without any resistance.
This is an important ratio that relates our known electrochemical reaction data $\bar{i}$, the nondegraded current, with the real volumetric current $R$.
In physical terms, $W$ can also be thought of as an effectiveness parameter that indicates the ability to accept current, representing the performance of a battery, since a higher effectiveness indicates a better performing particle.
However, a better performing particle is also more sensitive to degradation effects, so infinitely small particles are not the most beneficial in electrode design.

Intercalation currents in battery particles are usually modeled as surface reactions \cite{fuller1994simulation, smith2017multiphase}.
However, the total capacity in the system depends on the particle volume.
Solid diffusion is often not limiting in intercalation materials, where reaction limitations tend to be more important in nanoscale systems \cite{fraggedakis2020scaling}.
Under this assumption, there is a simple scaling of the particle size (radius $r$) relating the reaction rate per volume $R$ to the reaction rate per area $i$ as $V R = Ai$, where $V$ is the particle volume and $A$ is the particle area. 
Using this, for spherical particles, the fitness can be simplified to 
\begin{equation}
\label{eq:fitness_value}
W = \frac{1}{r} \frac{i}{\bar{i}},
\end{equation}
which relates the particle filling rate to the current density as shown earlier.
This introduces a separation between particle size effects $1/r$ and degraded charge transfer effects $i/\bar{i}$.
The initial distribution of the ``fitness'' is determined by the size of the particle distribution as $ W = 1/r$ since there is no degradation.

\subsection{Degradation Models}

From the degraded charge transfer effect of fitness $i/\bar{i}$, we see that the degradation buildup on each particle also plays a role in the fitness.
Mechanically and electrochemically, there are many different modes of degradation in a lithium-ion battery \cite{kabir2017degradation}.
For simplicity, we only consider the most important electrochemical degradation modes in lithium ion batteries.
There are three possible modes of electrochemical active degradation in a battery material that we will consider.
The first is formation of a film resistance on the battery, based on some film-forming reaction, such as solid-electrolyte interphase formation \cite{pinson2012theory, li2019investigations}.
The result of this is an increased surface resistance that mainly plays a role in affecting the kinetics by reducing the overpotential.

The second mode of degradation is through the reduction of active material, from phase transformations into rock salt or spinel phases (such as in nickel rich materials), which cause changes in surface kinetics and active material capacity \cite{mohanty2016modification, li2020direct}.
This would result in the loss of active material, which plays a role through rescaling the available lithium concentration in the system.
It mainly reduces the number of available sites in the transition state, which affects the reaction kinetics.

The last mode of degradation is a general contribution from degradation in the battery, either in the cathode or anode, as a loss of electrolyte in the system from degradation reactions.
This can be modeled with a loss in electrolyte concentration \cite{pinson2012theory, sarasketa2015understanding}.

Since we use nickel-rich materials as an example, degradation reactions are modeled at higher voltages \cite{li2020degradation, manthiram2016nickel} using a simple Tafel reaction to define the degradation current as
\begin{equation}
\label{ref:eq:deg}
    i_{deg} = k_{0,deg} \exp{\left({\mu_{deg}^0 - \mu_{res} + iR_f}\right)}
\end{equation}
where $\mu_{deg}^0$ is the cutoff potential for degradation.

\subsubsection{Resistive Film} \label{sec:resistive_film}

A resistive film may form as a type of solid electrolyte interphase on the cathode \cite{edstroem2004cathode, maleki2019controllable} or anode \cite{edstrom2006new} and grow continuously. 
Experimental measurements of resistive interface growth have found that though the initial amount of growth is quite rapid, even past the initial stages, there is often continuous growth of resistive film on active material \cite{huang2019evolution}.
For any intercalation reaction, when a resistive film grows, it affects the current $i$ through the reduction of the overpotential to $\eta + iR_{f}$ with film resistance $R_{f}$.

From the definition of the fitness of the reaction rate in Eq. \ref{eq:fitness_value}, it is necessary to obtain $i/\bar{i}$ when there is a resistive film.
The simplest electrochemical reaction is a symmetric Butler-Volmer model, which can be solved with fewer simplifications than the generic Butler-Volmer model because of the symmetry of the model.
If we do a Taylor expansion on this system and assume that the charge transfer coefficient is symmetric, we can obtain an analytic expression for the driving force term in Eq. \ref{eq:sym_BV_expand}.
We obtain the fitness value any given time as 
\begin{equation}
\begin{split}
   \label{eq:fitness_value1} W  = \frac{1}{r}\left[1 + \alpha i \coth{(\alpha \eta)} R_f\right] + \mathcal{O}(R_f^2).
\end{split}
\end{equation}

A more general case can be found by linearizing the full kinetic model with respect to the overpotential.
Using this, we can avoid the need to find an implicit solution of this problem.
A natural dependence on the \change{differential conductance} occurs from how resistance affects overpotential.
The ratio of the degraded to nondegraded current is found to be 
\begin{equation}
    \frac{i}{\bar{i}} = \frac{1}{1- R_f\frac{\partial\bar{i}}{\partial \eta} } + \mathcal{O}(R_f^2) 
\end{equation}
from the linearizations in Appendix \ref{ref:partial_derivs}1, where the second order solution can be seen in Appendix \ref{appdx:second_order}.
Thus, the fitness value is found to be
\begin{equation}
\label{eq:approximation1}
    W  = \frac{1}{r} \left(1-R_f \frac{\partial \bar{i}}{\partial \eta}  \right)^{-1} + \mathcal{O}(R_f^2),
\end{equation}
which is the general formula for any reaction rate.
The specific analytical formulas for each reaction rate (Butler-Volmer and CIET) can be found in Appendix A.

For the resistive film model, with the electrochemical cycling of a battery, the resistance on a particle $R_{f}~ (\Omega 
\cdot m^2)$ changes as
\begin{equation}
   \label{eq:resistance} \frac{dR_{f}(t,c, r)}{dt} = \beta  i_{deg},
\end{equation}
with the value of the degradation current per area $i_{deg}$ from Eq. \ref{ref:eq:deg}.
\change{The resistivity per amount of reaction is $\beta = n\text{MM}/(F\sigma \rho)$, which is the resistivity per amount of resistance reaction in units of $\Omega \cdot m^4/C$.
This is a property of the material which makes up the resistive film.
It is related to the conductivity of the material $\sigma$, the density of the material $\rho$, the number of lithium atoms per chemical formula in the film material $n$, Faraday's constant $F$, as well as the molecular mass of the film material MM.}
For materials such as lithium carbonate which form inorganic films in batteries, we expect the resistivity parameter $\beta$ to be on the range of $10^{-7}-10^{-5}~\SI{}{\Omega \cdot m^4/C}$ \cite{omar2016electrical}.
Since we are only concerned with the dependence of the resistance with the particle size, we can take the mean value of Eq. \ref{eq:resistance} over the concentration distribution using Appendix \ref{sec:avg_FP}. 

\subsubsection{Surface Blockage} \label{sec:surface_blockage}

The model of fitness in surface blockage is similarly defined to the approximate solution of the resistance formation model.
The surface blockage model is a homogeneous version of a model for phase transitions from cation disorder-induced degradation, especially common in nickel rich materials, which involves a change in surface concentration as well as bulk availability \cite{zhuang2022theory, lin2014surface, yan2017atomic}.
The rescaled capacity is defined as $\tilde{c}$.
Since there is a loss of active material in this model, in the chemical potential model of the active material, the real concentration needs to be rescaled by the amount of capacity loss as $\mu_c(c/\tilde{c})$ instead of $\mu_c(c)$.
The value of degraded to nondegraded current is shifted as a result, as the reaction is affected by the rescaled chemical potential as well as the reduction in the number of available sites, which influences the reaction rate through transition state theory.
We can calculate these two effects separately.

We first calculate the effects from the reaction rate without prefactor, which comes from the effect of overpotential on this mechanism.
The ratio of the degraded to nondegraded reaction rate without the prefactor is found to be 
\begin{equation}
    \frac{h}{\bar{h}} \approx 1 + \frac{1}{\bar{h}} \frac{c(1-\tilde{c})}{\tilde{c}^2} \frac{\partial \mu_c}{\partial c} \frac{\partial \bar{h}}{\partial \eta}
\end{equation}
from a Taylor expansion of $h$ in Appendix \ref{ref:partial_derivs}2, where we again see a form similar to the \change{differential conductance}.
This change in maximum capacity plays a role by limiting the current in these coupled-ion electron transfer reactions.
If we mainly consider surface effects, we can neglect this term.

Following, we can calculate the effects of the prefactor ratios, which consist of the surface effects.
For an nondegraded Butler-Volmer current, the ratios of the prefactors is $\left(\frac{\tilde{c}-c}{1-c}\right)^{1-\alpha}$ from the thermodynamically consistent Butler-Volmer equation \cite{bazant2013theory}.
For the coupled-ion electron transfer reaction, the ratios of the prefactors is $\frac{\tilde{c}-c}{1-c}$ when the transition state is assumed to occupy one site.
Thus, the total fitness for the surface blockage model results in
\begin{equation}
    W \approx \frac{1}{r} \left(\frac{\tilde{c}-c}{1-c} \right)^n + \mathcal{O}((\tilde{c}-1)^2),
\end{equation}
where $n = 1-\alpha$ for the Butler-Volmer equation, and $n = 1$ for the coupled-ion electron transfer reaction rate.

For a model of surface blockage, the amount of degradation is classified by 
\begin{equation}
    \frac{d\tilde{c}(t,c,r)}{dt} = -\frac{i_{deg}C}{\rho_{s,max} r}
\end{equation}
an equation that scales with the size of the battery particle and the total site density of the material.
Here, $C$ is the Coulomb number and $\rho_{s,max}$ is the maximum site density of the material in mol/m$^3$.
This equation can again be integrated over all concentration values for an average value per particle size as in Appendix \ref{sec:avg_FP}.


\subsubsection{Electrolyte Loss} \label{sec:loss_lithium_inventory}

In a full battery cell, the amount of degradation should be affected by degradation on the other electrode (anode) as well. 
The formation of degradation on the anode will often lead to a loss of usable lithium capacity from the lithium consumed in the side reaction to form products \cite{zheng2015correlation}.
These products consist of the solid electrolyte interphase, cathode electrolyte interphase, and others.
\change{Since we are modeling a perfect electrolyte bath, which does not use an opposing electrode}, we cannot ``consume'' lithium ions on the other electrode and reduce the total usable lithium concentration.
Thus, \change{electrolyte loss} needs to be prescribed in the system, which we choose to be linear for simplicity \cite{pinson2012theory}.
\change{We call this degradation mechanism \change{electrolyte loss} since it reduces the available electrolyte.}

In the Butler-Volmer equation, the electrolyte concentration affects the transition state prefactor as $k_0(c) \propto c_+^{1-\alpha}$.
In addition, there is a subtle effect on shifting the overpotential from the entropic component.
From linearizing the reaction rate without prefactor $h$ in Appendix \ref{ref:partial_derivs}3 and applying the definition of the thermodynamic factor $\frac{\partial \ln{a_+}}{\partial \ln{c_+}}$ from Ref. \cite{newman2012electrochemical}, which relates the activity in a Stefan-Maxwell concentrated electrolyte with the lithium ion concentration, we can obtain the fitness value.
We see that the fitness value in a Butler-Volmer equation is simply found to be 
\begin{equation}
    W = \frac{a_+^{1-\alpha}}{r}\left[1 +\frac{1}{\bar{h}} \frac{\partial \ln{a_+}}{\partial \ln{c_+}} k_B T\frac{\partial \bar{h}}{\partial \eta} (1-c_+) \right] + \mathcal{O}((c_+-1)^2)
\end{equation}
with our linear approximation for concentrated solutions.
A dilute solution approximation can also be used, where the thermodynamic factor is unity.

For coupled ion electron transfer, the effect of electrolyte is more complicated \cite{Zhang2022_CIET_preprint}.
In the reduction reaction, since the electrolyte is a reactant, there is a direct concentration dependence on the reduction reaction, but not on the oxidation reaction.
In addition, in the formal overpotential, the electrolyte does not influence the amount of available sites in the overpotential except through the activity in a concentrated solution.
Thus, it does not change the overpotential of the reaction for a dilute model \cite{fraggedakis2021theory}.
\change{
As derived in Appendix \ref{ref:partial_derivs}3, we see that 
\begin{equation}
    W = \frac{1}{r}\left[1-  \frac{\partial \ln{a_+}}{\partial \ln{c_+}}\frac{\bar{i}_{red}}{\bar{i}}\left(1-c_+\right)\right] + \mathcal{O}((c_+-1)^2)
\end{equation}
is the fitness value for the \change{electrolyte loss} model if a CIET reaction rate is used.}
For a CIET reaction rate, the reduction current ratio in the total reaction contributes strongly to scaling the value of the fitness function.
This causes the fitness in intercalating systems to be lower than in deintercalating systems for this model.

Assuming that the initial electrolyte concentration is unity, the degradation rate for a half-cell model with a \change{electrolyte loss} can be prescribed using the simple relation
\begin{align}
c_+ = 1-kt,
\end{align}
where \change{$k>0$ is a parameter that reduces the availability of electrolyte with time in units of M/hr if the initial concentration is 1 M} and $t$ is the amount of time spent cycling the battery in hours.

\subsubsection{Combined Model}
These three degradation models can be combined into an overall fitness value to account for multiple degradation effects to the first order approximation.
Similarly, if other fitness values for different degradation mechanisms are also derived, they can be combined into such an overall fitness value.
The value of the combined fitness function for the Butler-Volmer equation is 
\begin{widetext}
\begin{equation}
\label{eq:W_BV}
    W \approx \frac{a_+^{1-\alpha}}{r} \left( \frac{\tilde{c}-c}{1-c}\right)^{1-\alpha} \left[1 + \alpha i \coth({\alpha \eta}) R_f\right]  \left[1 +\frac{1}{\bar{h}} \frac{\partial \ln{a_+}}{\partial \ln{c_+}} \frac{k_B T}{e}\frac{\partial \bar{h}}{\partial \eta} (1-c_+) \right].
\end{equation}
\end{widetext}
For the coupled-ion electron transfer system, the overall fitness function can be written as 
\begin{widetext}
\begin{equation}
\begin{split}
\label{eq:W_CIET}
    W & \approx \frac{1}{r} \left(\frac{\tilde{c}-c}{1-c} \right) \frac{1}{1- R_f \frac{\partial \bar{i}}{\partial \eta}} \left[1-\frac{\bar{i}_{red}}{\bar{i}}\left(1-c_+\right) \frac{\partial \ln{a_+}}{\partial \ln{c_+}}\right].
\end{split}
\end{equation}

\end{widetext}

From each of these equations, we can see the explicit contributions of particle size and the three different degradation modes we are modeling (resistive film, surface blockage, and electrolyte loss).
The separate effects of each degradation mode contribute to the overall fitness, aiding understanding of which modes are the most detrimental and should be mitigated to preserve the current capability of the battery.

\section{Simulations}

\subsection{Numerical Setup}

Here, we attempt to model the degradation of a \change{nickel manganese cobalt oxide blend electrode with a ratio of 5:3:2 (NMC532) under constant current cycling} and capture the evolution of capacity changes and degradation with time evolution.
Using the Fokker-Planck model in Eq. \ref{eq:approx_FP}, we model a single electrode volume.
\change{We simulate each degradation mode separately to analyze their individual effects.}
Simulations were performed with MATLAB \change{using autodifferentiation} from CasADi \cite{andersson2018} to increase speed of solving the DAE system.
The Fokker-Planck numerical simulations were performed until end of life for each degradation and reaction model.
Simulation parameters and details were reported in Appendix \ref{ref:appdx_sim_params}.

\subsection{Analysis}

\begin{figure*}[t]
\includegraphics[width=0.5\textwidth]{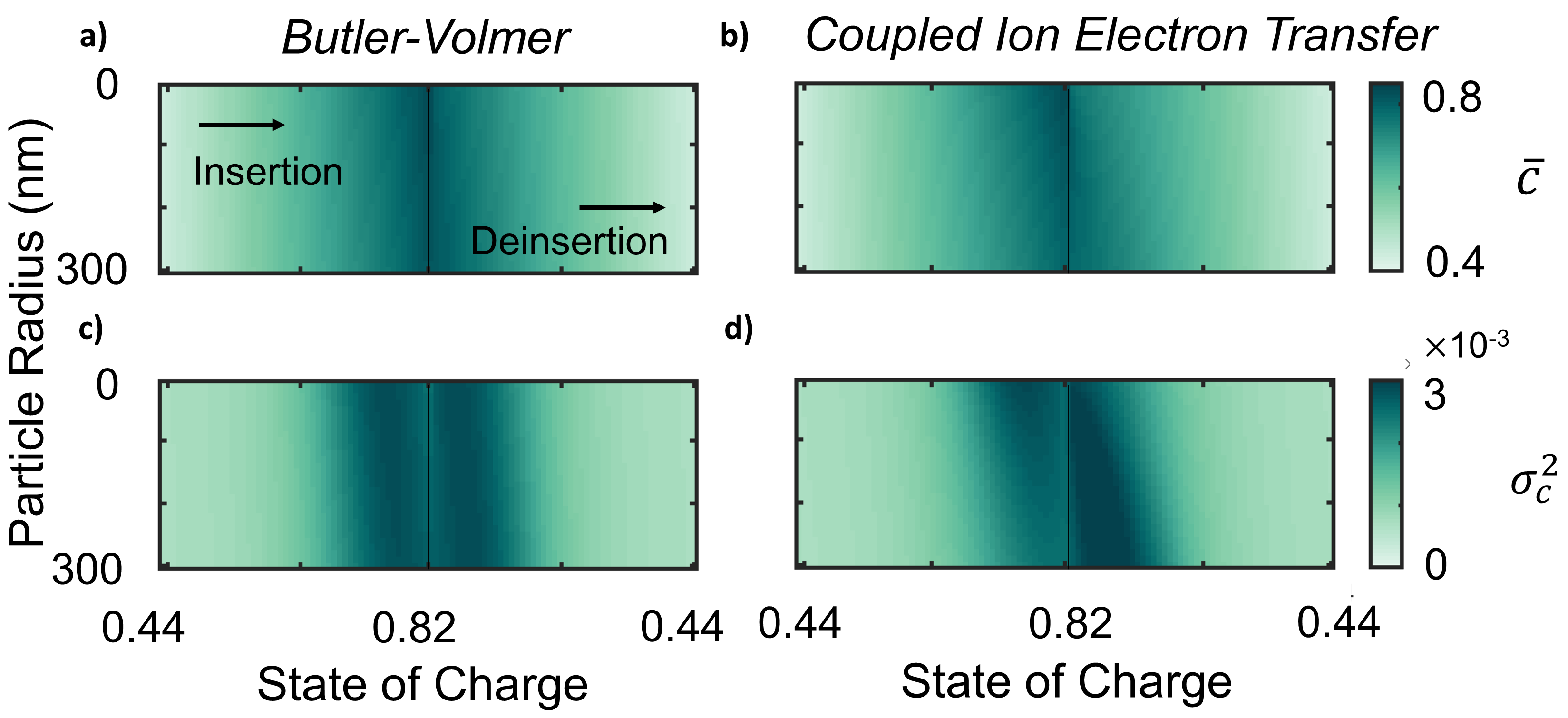}
\centering
\caption{a,b) Average solid lithium concentration for a single cycle of charging and discharging for Butler-Volmer/coupled ion electron transfer plotted with respect to the state of charge and the particle size during charge and discharge. c,d) Variance of the solid lithium concentration at a single cycle for each state of charge plotted with respect to the state of charge and the particle size from charge to discharge.}
\label{fig:fig1}
\end{figure*}

We first focus on the particle level details of degradation, and then analyze how dynamics at the microscale affects electrode-level degradation.
From the particle level details of kinetics and degradation, we observe the heterogeneity between particles in intercalation.
To understand their contributions to degradation, the fitness values in the simulation are observed.
The heterogeneity of degradation at the particle level scale and the autocatalytic relationship between fitness and degradation are discussed.
Following this, microscale degradation is then used to explain electrode-level phenomena, with the voltage curves and the capacity loss data as an example.
Heterogeneity at the particle level is then found to heavily influence electrode-level degradation effects, especially from the smaller particles.

\subsubsection{Particle Level Heterogeneities}

In a single cycle, the average concentration at each particle radius is plotted in Fig. \ref{fig:fig1}a,b and the variance of the concentrations is plotted below in Fig. \ref{fig:fig1}c,d.
We observe that even the differences between these reaction rates can cause heterogeneity in the intercalation of particles.
This is influenced by the ``limiting current'' for coupled-ion electron transfer type reactions from the electron availability, while the Butler-Volmer reaction grows exponentially with overpotential and is unbounded.
This limiting current is bounded by the transition state value $\gamma^\ddag = (1-c)^{-1}$, which shows up in both reaction rates, but because of the exponential form of the Butler-Volmer equation, it affects the coupled-ion electron transfer reactions more strongly.
This causes an asymmetry between charge and discharge in the CIET reaction rate, stemming from the prefactor.
The asymmetry in discharge is discussed further in Refs. \cite{zhao2019population, park2021fictitious}.

\begin{figure*}[t]
\includegraphics[width=0.5\textwidth]{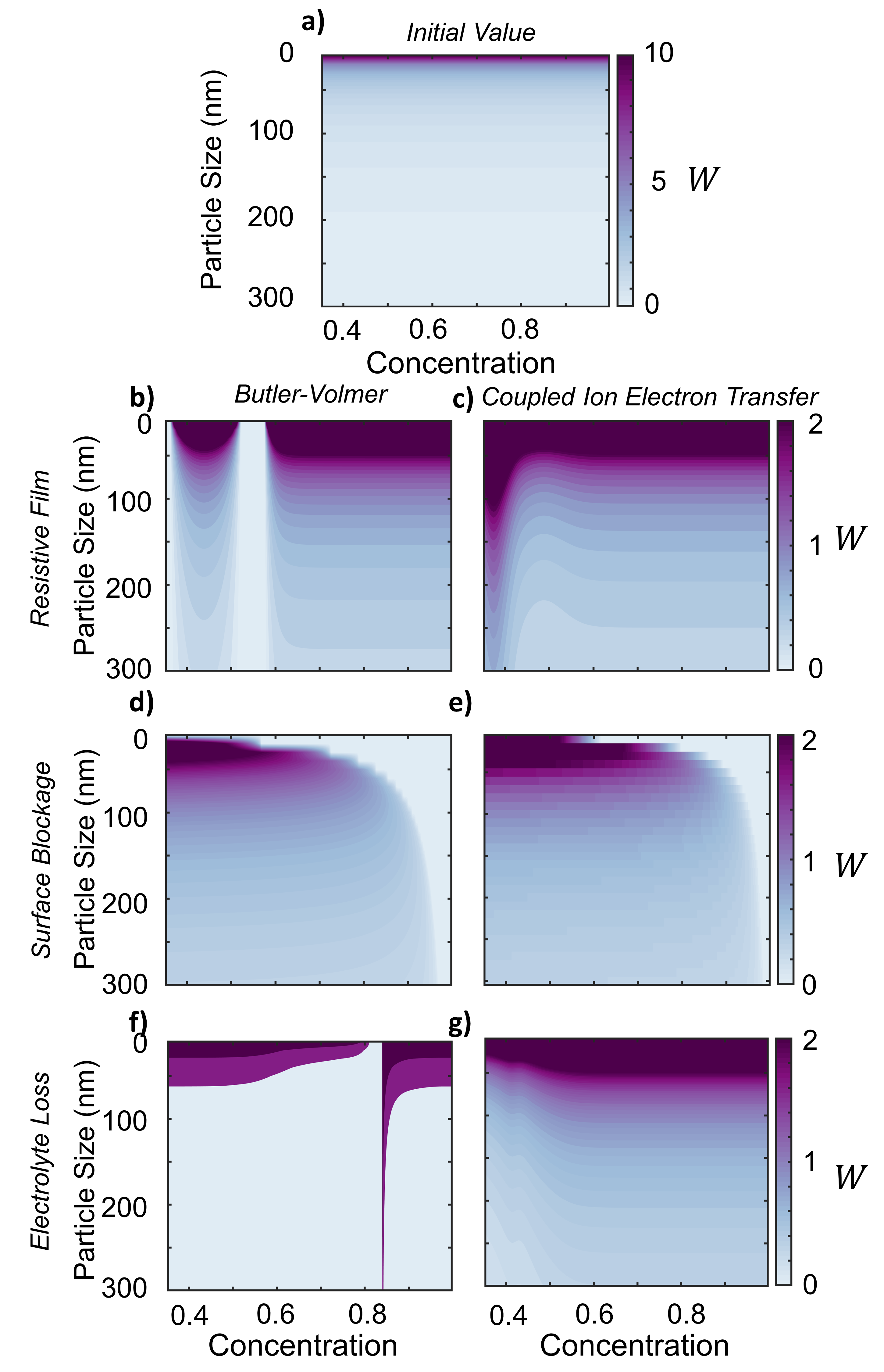}
\centering
\caption{a) The initial fitness value $\bar{W} = 1/r$ is plotted in the first cycle with respect to each concentration and particle size at a single time point during the initial cycle.
b,c,d,e,f,g) For each Butler-Volmer/coupled ion electron transfer reaction rate, the fitness values $W$ from a set of simulations each with a single degradation mode towards the end of cycling are plotted with respect to concentration and particle size at a single time point during one of the last few cycles.}
\label{fig:fig4}
\end{figure*}

The heterogeneity at the particle level scale representing the effectiveness of each particle is described by the fitness.
This describes each particle's inherent current-carrying capability. 
The initial value of performance is defined by the inverse particle size as shown in Fig. \ref{fig:fig4}a.
After cycling, the fitness function $W$ is plotted in Fig. \ref{fig:fig4}, where the analytical values can be found in Appendices \ref{sec:resistive_film}, \ref{sec:surface_blockage}, and \ref{sec:loss_lithium_inventory}.
The fitness becomes infinitely small from the unbounded behavior of Butler Volmer reactions for some values in Fig. \ref{fig:fig4}b, where the reaction is infinitely large for intermediate concentrations.
This causes asymptotic behavior in the Butler-Volmer solution for the resistive film and \change{electrolyte loss} models.

\begin{figure*}[t]
\includegraphics[width=0.5\textwidth]{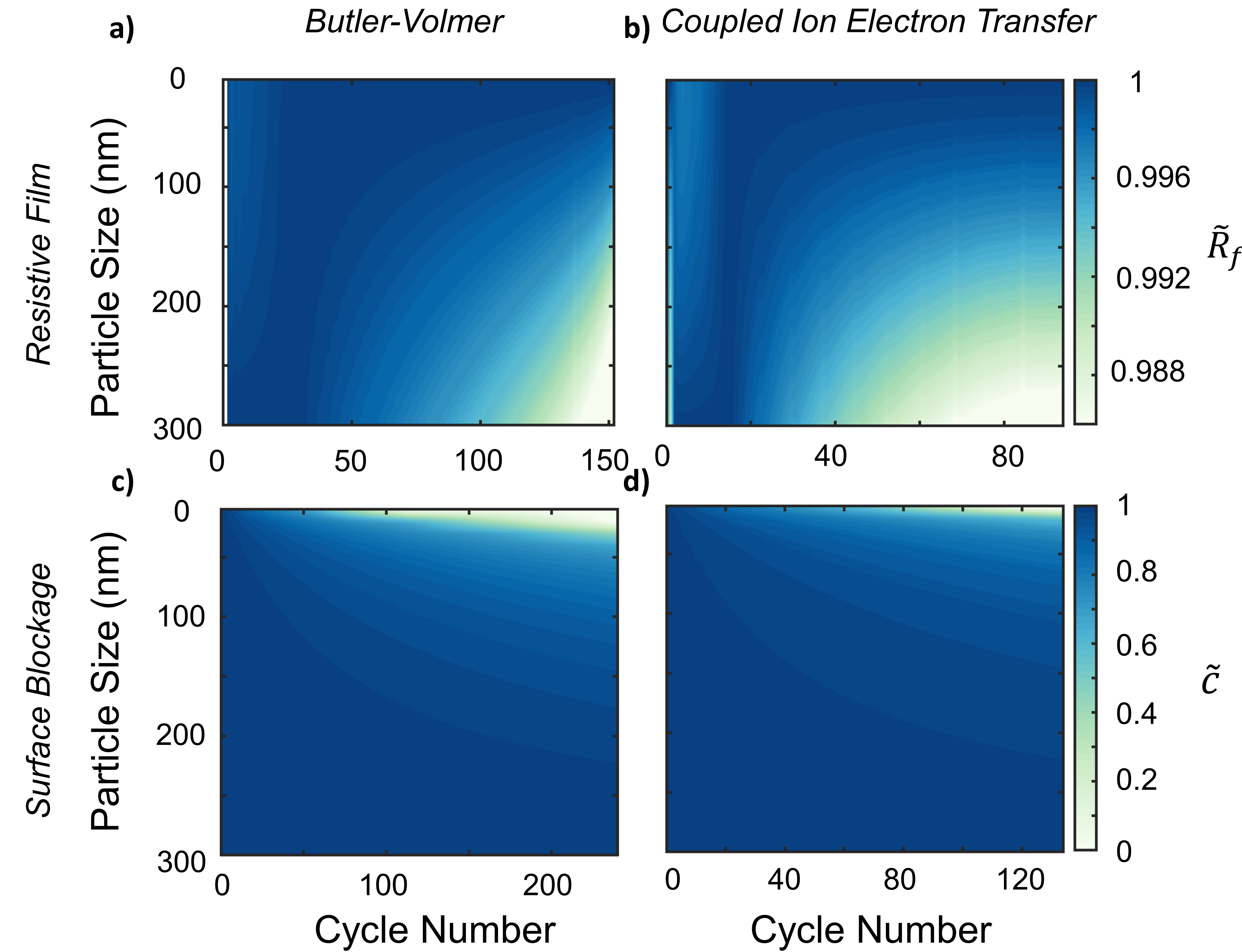}
\centering
\caption{a,b) For a set of simulations with only resistive film growth, relative resistance at each cycle to the maximum resistance at each cycle is plotted with respect to cycle number and particle size, for the Butler-Volmer and coupled ion electron transfer reactions.
c,d) For a set of simulations with only surface blockage increase, surface blockage at each cycle relative to the maximum surface blockage at each cycle is plotted with respect to cycle number and particle size, for the Butler-Volmer and coupled ion electron transfer reactions.
(The degradation mechanism of \change{electrolyte loss} is prescribed so it has no heterogeneity). This plot displays the heterogeneity growth in degradation as we cycle the battery.}
\label{fig:fig2}
\end{figure*}

We observe that as the battery is cycled, the value of the fitness function is reduced from degradation accumulation.
Overall, the reduction of the fitness values drives further increases in heterogeneity in degradation, which triggers more heterogeneous reduction in the fitness values.
Thus, we observe an autocatalytic effect between fitness and degradation.

This autocatalytic relationship between fitness and degradation is further seen in Fig. \ref{fig:fig2}.
The relative degradation parameter with respect to the maximum parameter at each time is plotted for the Butler-Volmer or coupled ion electron transfer reactions with the degradation mechanisms from film resistance or surface blockage.
Resistance growth asymmetry happens in the first couple cycles in Fig. \ref{fig:fig2}a and b, where more resistance forms on the larger particles and reaches steady values after the first few cycles to attempt to homogenize the system.
Initially, no autocatalytic behavior is observed as the initial resistance formation seen is part of the ``formation cycling'' in battery electrodes \cite{an2016state, an2017fast} to stabilize the system and reduce the homogeneity between the particle sizes.

The second step of nonheterogeneity in resistance formation appears after formation cycling in the battery lifetime and reduces the overall capacity of the battery as shown in Fig. \ref{fig:fig2}.
Because of the larger fitness values for small particles, there is a higher capacity change fraction distributed to them, causing more degradation from the higher amounts of degradation current.
This degradation growth behavior becomes autocatalytic as more asymmetry in degradation growth is observed with cycling.
This is seen in the later cycles in Fig. \ref{fig:fig2}ab, where more resistance forms on the smaller particles towards the end of life.
Similarly, for the surface blockage mechanism, there is an autocatalytic effect on degradation favoring smaller particles.
This is coupled with the fitness values, which eventually lose all their available capacity towards end of life and leaves only larger particles operational.
Thus, for both the surface blockage and resistance formation models, heterogeneity grows autocatalytically as we cycle the batteries, with degradation favoring the smaller particles and larger particles remaining more stable.

From the combined observations of particle level degradation at different particle sizes, we theorize that the terminating behavior of battery capacity is not caused by the average particle size, but rather the smaller particle sizes.
As observed in Fig. \ref{fig:fig2}, at later stages in cycling, degradation starts to accumulate on all particles, but especially quickly on smaller particles, which require larger amounts of potential to charge/discharge and thus causes a stronger drop in battery capacity.
This autocatalytic behavior between fitness and degradation drives a strong heterogeneity in fitness values in Fig. \ref{fig:fig4} as the battery is cycled.
This continually favors smaller particles as the system reaches end of life in Fig. \ref{fig:fig2}.

\subsubsection{Electrode-Level Measurements}

\begin{figure*}[t]
\includegraphics[width=0.5\textwidth]{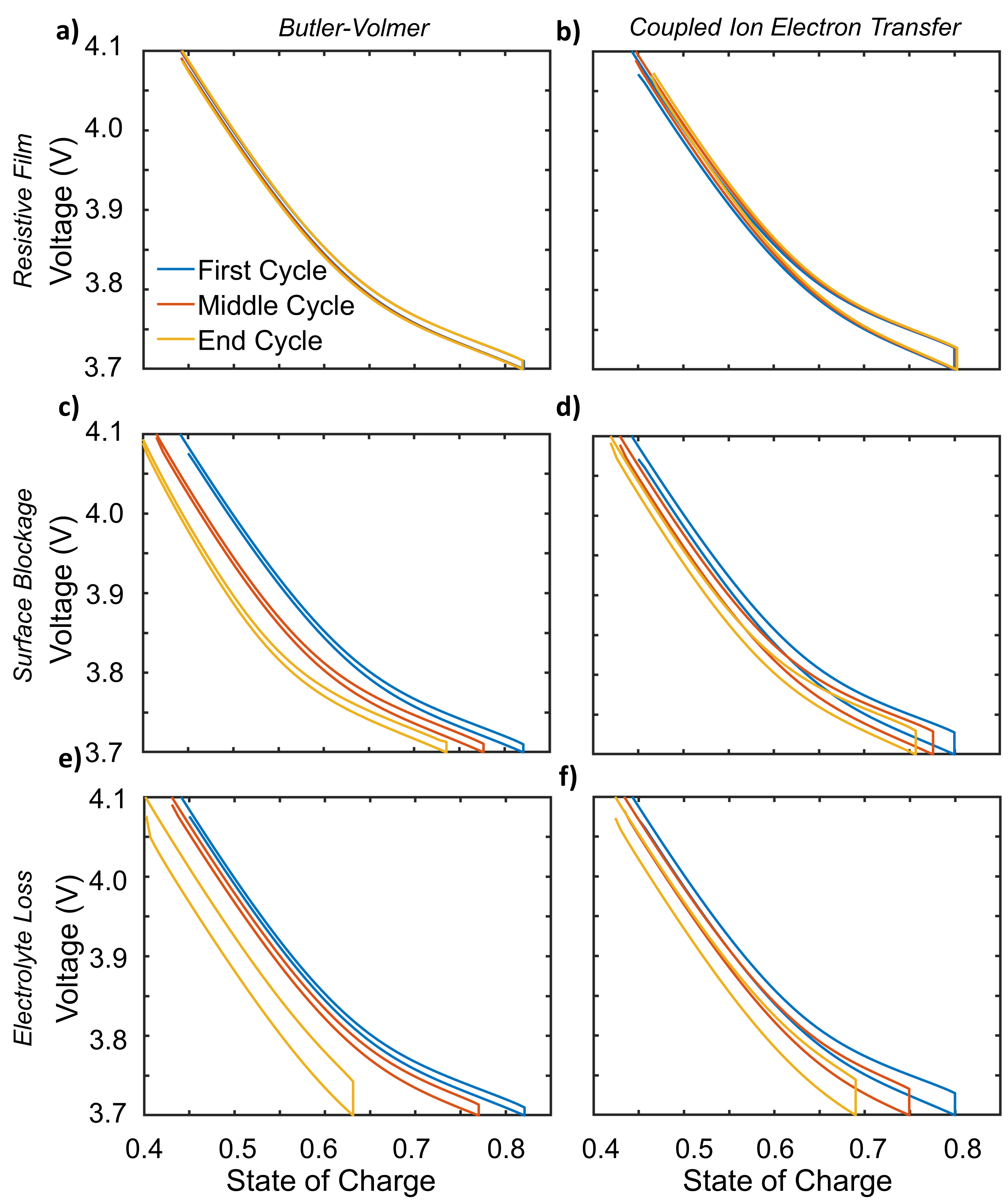}
\centering
\caption{The voltage discharge and charge curves from simulations with a single degradation mechanism with respect to the state of charge of the cathode are plotted at the beginning, middle, and end cycle of each set of simulations with each reaction model (BV/CIET for a,c,e and b,d,f) and degradation mechanism (resistive film for a and b, capacity loss for c and d, and \change{electrolyte loss} for e and f).}
\label{fig:fig6}
\end{figure*}

The effects of heterogeneous degradation effects can be shown to influence the experimental measurements of electrode-level electrochemical phenomena.
The shifting and hysteresis behavior of the voltage curve provides important pieces of electrode-level information from cycling the degraded cells.
The capacity loss profile can also be observed and information on the degradation mechanisms can be extracted from the shape of these curves.

Expansion of the voltage curves, or hysteresis, occurs from kinetic limitations in the system, which are often caused by degradation mechanisms that do not degrade the active material.
For the \change{electrolyte loss} mechanism in Fig. \ref{fig:fig6}, the formal overpotential for the coupled-ion electron transfer reaction does not contain any electrolyte effects.
The main influence of degradation is on the kinetics in the reduction direction of the reaction.
Thus, there is no shift in the open circuit voltage (OCV) curves and only an ``expansion'' of the charge/discharge curve around the original open circuit voltage, which occurs from the limitations on the kinetics.
The apparent shift in the open circuit voltage comes from the need to apply higher electrolyte potentials to compensate for lower electrolyte concentrations.
There is an asymmetry between insertion and deinsertion since the influence of electrolyte concentration appears solely in the reduction reaction.

Shifts in the voltage measurement, however, generally occur from either shifts in overpotentials required or active material ranges.
The surface blockage mechanism observes a leftward shift compared to the potential of the original model, occurring from the reduction of available transition state sites and the shift in the OCV.
This causes the change in active material range in the surface blockage mechanism, which generates a shift in the overpotential ranges, translating to the accessible voltage range.
Similarly, because of the shifted overpotentials for the Butler-Volmer formulation for the \change{electrolyte loss} mechanism, there is a downward shift of the open circuit voltage curve.
This is in contrast to the surface blockage mechanism, since there is not a change in range of active material voltage, but a shift in the electrolyte potential applied.
This generates a downward movement on the voltage curve instead of leftward shift of charging range.

The electrode-level behavior of ``expansion'' and ``shifting'' of the discharge curves provides us with macroscopic information on the contributions of the degradation mechanisms from kinetic effects or changes in the overpotential (which can be caused by electrolyte concentration loss or active material degradation).
Generally, a combination of these will contribute to the physical degradation of voltage curves.
Through observing the expansions and shifts of the degraded charge/discharge curve, we can learn about whether degradation consists of active material degradation or electrochemical changes in kinetics at the surface.

\begin{figure*}[t]
\includegraphics[width=\textwidth]{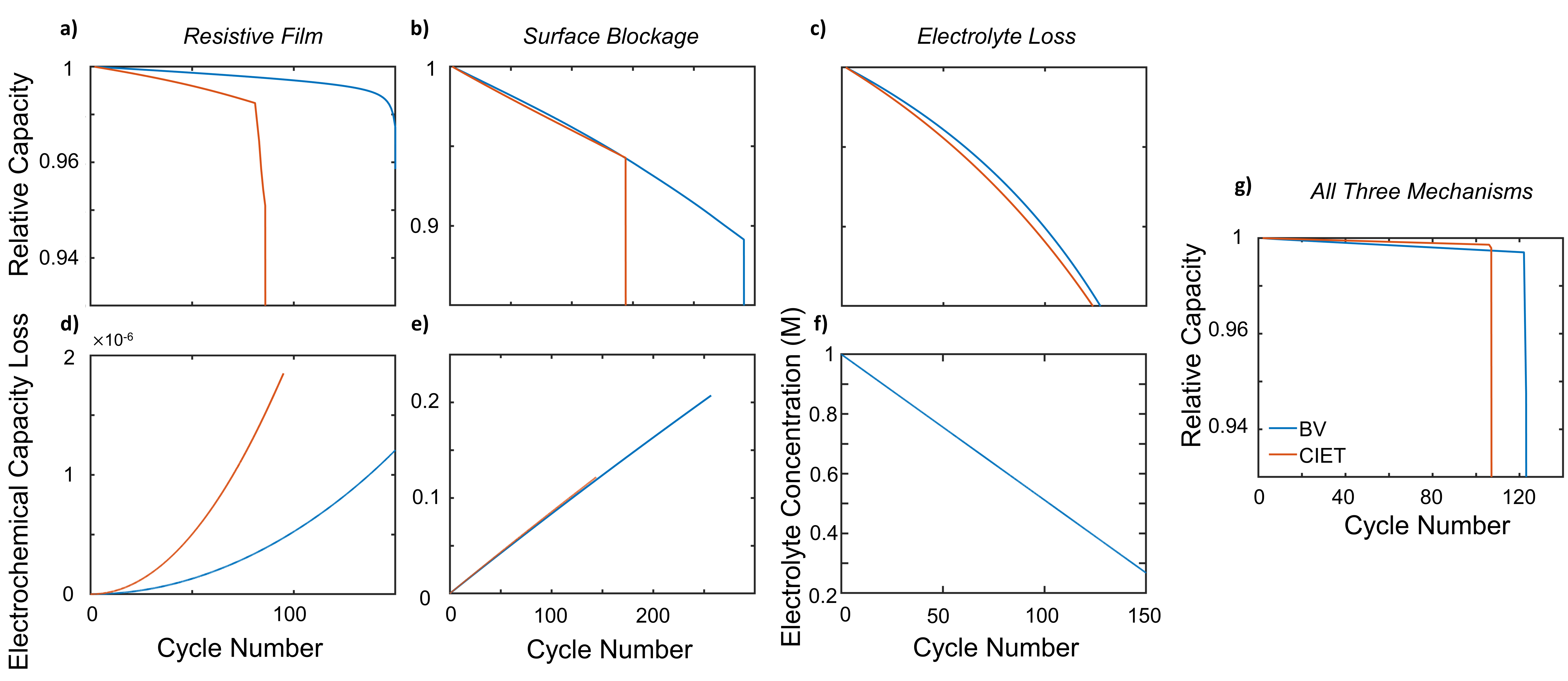}
\centering
\caption{a,b,c) The capacity cutoff for each cycle with respect to the cycle number for BV/CIET is plotted for sets of simulations with the three separate degradation mechanisms in the first row, while the real relative electrochemical capacity loss from the integrated degradation current with respect to the cycle number is plotted in the second row in d) and e). For the \change{electrolyte loss} model, because the electrochemical capacity loss is from the anode, we instead plot the prescribed degradation of electrolyte concentration with respect to cycle number in f). g) We plot the capacity loss curve with respect to cycle number for a set of simulations with all three degradation mechanisms implemented for the two reaction models.}
\label{fig:fig5}
\end{figure*}

In capacity loss curves, a linear drop in capacity is observed in Fig. \ref{fig:fig5}, which later rolls over into a steeper capacity loss curve when more degradation has accumulated.
Each degradation mechanism contributes differently to the terminal behavior of the capacity.
The resistance formation mechanism and surface blockage mechanism both cause sharp terminating curves, but the resistive film mechanism is much smoother than that of the surface blockage mechanism.
The \change{electrolyte loss} mechanism has a smooth and nonlinear drop off as the concentration of electrolyte drops, especially at lower electrolyte concentrations.

The different shapes of these drop offs may be able to give insight into the dominance of different degradation mechanisms from the experimental observation of capacity loss plots.
In addition, a set of simulations where all three models were combined was performed, where it can be seen that the dominant degradation mechanism tends to override the capacity loss curve (in, this case, the surface blockage mechanism) in Fig. \ref{fig:fig5}g.


\section{Conclusion}

Using the idea of fitness functions in biological populations, we map this idea to the degradation of battery particles of varying sizes in a lithium-ion battery.
The coevolution of fitness with reaction and degradation on these battery particles is modeled as the battery is cycled in the Fokker-Planck equation, using different intercalation reaction models.
For all reaction models, the fitness function values are found to grow heterogeneously as the battery is cycled, initially with formation cycles accumulated onto the battery.
After the formation cycles, degradation accumulates while preferring smaller particles.
The observed heterogeneous degradation accumulation on the smaller particles from the autocatalytic behavior between fitness and degradation contributes to the eventual death of the battery \cite{attia2022knees}.
This overall trend causes the smaller particles to lose their usable capacity faster and contribute to the failure of the battery before the larger particles do.

From our simulations, we also learn that particle level degradation mechanisms drive electrode-level behavior of the system, and the shape of the behavior of electrode-level current-voltage relations gives insight into which of the degradation mechanisms is most dominant.
``Expansion'' of voltage curves can be attributed to changes in reaction kinetics, while ``shifts'' of the voltage curves are attributed to active material degradation or changes in overpotential. 
Asymmetric effects between intercalation and deintercalation in the voltage curve can be attributed to \change{electrolyte loss} effects from their stronger contribution to the reduction reaction.
In addition, the shape of the capacity loss curve gives insight into the mode of degradation that is most dominant.
Electrolyte loss has smoother terminal behavior, while resistive film and surface blockage all have sharper capacity loss drops.

\change{Future work would extend to experimental validation of this model. 
Possible experiments to perform could involve SEM or TEM imaging experiments on observing degradation growth on particles from the film thickness dependent on particle size \cite{sacci2014direct, huang2019evolution}.
Small and large particle distributions could also be mixed to make observations of degradation growth simpler, so that only two limits of degradation values would need to be observed and compared.}

Such a model of fitness evolution for driven electrochemical reactions can be expanded to systems beyond a simple Fokker-Planck model.
These simple degradation mechanisms can be applied to porous electrode models \cite{doyle1993modeling, fuller1994simulation, newman2012electrochemical} to study the effect of degradation on the porous electrode scale.
The development of simple, physically driven degradation models which can be applied to porous electrode scale simulations can provide support for data-driven modeling of degradation, aiding solutions to the major challenges in developing and designing better Li-ion batteries \cite{aykol2021perspective}.

\begin{acknowledgments}
The authors acknowledge Marc D. Berliner for help in modifying the code to decrease the computational time for simulations as well as Huada Lian, Dimitrios Fraggedakis, Mehran Kardar, Huanhuan Tian, and Kranthi K. Mandadapu for insightful discussions and help formulating the manuscript.
The authors also acknowledge Xiao Cui for input on experimental procedures, and Akiva G. Gordon for help editing the manuscript.
This work was supported by the Toyota Research Institute through D3BATT: Center for Data-Driven Design of Li-Ion Batteries.

\end{acknowledgments}

\begin{widetext}

\clearpage

\appendix

\section{Derivation of the Modified Fokker-Planck Equation} \label{ref:appdx_derivation_FP}
The derivation of Fokker-Planck with resistance evolution in these systems is similar to that of biological evolution with a fitness landscape. 
In fact, we expect to see the fitness effects more strongly as we apply current/voltage to a control system instead of letting the population evolve naturally as in genetics.
Starting from the Langevin equation, we can derive the Fokker-Planck equations needed as follows. 
The Langevin equation for the filling of a battery particle, assuming no solid diffusion limitation, is
\begin{equation}
     V_j \dot{c}_{j} = -  A_j\left(\change{i_j(\Delta \mu)} +  F_j(i_j)\right),~~j = 1\dots N,
 \end{equation}
where $V$ is the total volume of the particle, $A$ is the area of the particle, $c$ is the concentration of the particle, which is the site density of the intercalation material scaled by the maximum site density, $c = \rho/\rho_{s,max}$, $i$ is the reaction rate caused by the difference between the reservoir potential $\mu_{res}$ and the particle potential $\mu$, which is related to the driving force \change{$\Delta \mu = \mu_{res} - \mu$}, and $F(i)$ is the random force for particle $j$, which depends on the reaction magnitude $i$ by the fluctuation-dissipation theorem \cite{zwanzig2001nonequilibrium, kondepudi2014modern}.
In a battery particle, we know the volume of a particle scales with the particle size as $V_j \propto r_j^3$ and the surface area scales with the particle size as $A_j \propto r_j^2$, where $r_j$ is the radius of a spherical particle.
We can simplify the Langevin equation into 
\begin{equation}
    \dot{c}_{j} = -  \frac{1}{r_j}\left(i_j +  F_j(i_j)\right),~~j = 1\dots N.
\end{equation}
In a reactive system, for $j = 1 \dots N$ particles, every particle obeys the Langevin equation under a total constraint.
\change{Reactions happening directly between the particles are assumed to be nonexistent \cite{li2018fluid} and reactions only happen between the active material particles and the environment.}

This constraint is usually some form of constant current or voltage.
For constant current, we expect that $\sum_j m_j \dot{c}_{eff,j} = \text{Rxn}_{constraint}$ for applying a current constraint from all particles, or for constant voltage that $\mu_{res} = \mu_{constraint}$ for all particles $i$. 

If we assume that the probability density of particles in the system is at $f(c)$, we can convert this system of Langevin equations into a Fokker-Planck equation (acknowledging that the system is not deterministic from the noise term $F(i)$) \cite{risken1996fokker, frank2005nonlinear}.
\change{At each concentration variable, we need to take a small time increment such that we can sum the transition rates leaving and arriving at the current concentration as the chemical master equation
\begin{equation}
    f(t+\Delta t, c;r) = f(t, c;r) + \Delta t \int_0^1 p(c-c',c')f(t, c';r) W(t, c',r)dc' - \Delta t \int_0^1 p(c',c)f(t,c;r) W(t,c,r)dc'.
\end{equation}
Here, $p(x,y)$ is the transition matrix from state $y$ to $x$ before dampening the transition probabilities with the fitness.
It needs to be weighed by the changes in the transition state matrix because there is a modified amount of transitions happening from the changes in the fitness variable.}

The derivation of Fokker-Planck with resistance evolution in these systems is similar to that of a fitness landscape in genetics. 
We can start with the chemical master equation, which comes from the continuum limit of the previous equation to derive the full Fokker-Planck equation
\begin{equation}
    \frac{\partial f(t,c;r)}{\partial t} = \int^* dc' p(c-c', c')  f(t, c'; r)W(t,c',r) - \int^* dc' p(c'-c, c) f(t,c;r)W(t,c,r),
\end{equation}
and change the integration variable $c$ or $c'$ to the difference between these two variables $y = c - c'$.

\change{We take a Taylor expansion of the system only around the change of concentration variable $y$.
The Taylor expansion on the term in the integral returns
\begin{equation}
\begin{split}
    &p(y, c-y)f( c-y) W(c-y) = p(y, c)f(c)W(c) \\
    &- y \frac{\partial }{\partial c}\left( p(y, c)f( c)W(c)\right)\frac{y^2}{2}+\frac{\partial^2}{\partial x^2} \left( p(y, c)f( c)W(c)\right) + ...
\end{split}
\end{equation}
We know that there is no flux when there is no concentration change, so then the zeroth order term from the expansion is zero. Thus, only including the first and the second order term of the expansion, we see that the chemical master equation now is converted to the Fokker-Planck equation, which after moving the integrals into the derivative terms is revealed as
\begin{equation}
\begin{split}
    \frac{\partial f(t,c;r)}{\partial t} & = - \frac{\partial}{\partial c}\left[f(c)W(c)\left(\int dy y p(y, c) \right)\right] + \frac{1}{2} \frac{\partial^2}{\partial c^2}\left[f(c)W(c)\left(\int dy y^2 p(y, c) \right)\right]
    \end{split}
\end{equation}
to obtain the Fokker-Planck equation we are familiar with.
For simplicity in notation, we define the first order term as the reaction rate such that $i = \int_0^1 dc'c'p(c',c) $ to represent the driving velocity from the mean in concentration change from the transition probabilities.
The second order term as $D = \int_0^1 dc'c'^2p(c',c)$ represents the fluctuations, or the variance from the concentration change for the transition probabilities. 
From the definition of the intercalation reaction rate and the definition of a diffusion coefficient explicitly defined above, we see that
\begin{equation}
    \frac{\partial f}{\partial t} = - \frac{\partial}{\partial c}\left(fR\right) + \frac{\partial}{\partial c}\left(D \frac{\partial}{\partial c}\left(fW\right)\right). 
\end{equation}
where $D = k k_B T/N_t$ and $k = i/\eta$ for a linear form of the reaction rate mapping to previously electrochemical Fokker-Planck methods.
This is the full Fokker-Planck equation with the resistance evolution terms \change{\cite{zhao2019population, bazant2017thermodynamic, herrmann2014rate, dreyer2011behavior}}.}

\section{General Properties of the Fokker-Planck Model}
\label{sec:avg_FP}
Some general properties need to be defined in these systems.
Since we assume there cannot easily be particle size changes in the system, we see that 
\begin{equation}
    \int_0^1 f(t,c;r)dc = g(r),
\end{equation}
which is constant with time as the size distribution $g(r)$ of the particles.
The average over all the volumes for any property $x$ can be defined as
\begin{equation}
    \langle x\rangle = \frac{\int_0^\infty \int_0^1 xr^3 f(t,c;r) dcdr}{\int_0^\infty \int_0^1 r^3 f(t,c;r) dcdr} = \frac{\int_0^\infty \int_0^1 x r^3 f(t,c;r) dcdr}{\int_0^\infty r^3 g(r) dr},
\end{equation}
since the volume of each particle is of a different size, which scales with $r^3$.

\section{Second Order Solution to Resistive Film Model} \label{appdx:second_order}

The second order Taylor expansion to the resistive film model is as below:
\begin{equation}
    i = \bar{i} +\frac{\partial i}{\partial R_f} \bigg\lvert_{R_f = 0} R_f + \frac{1}{2}\frac{\partial^2 i}{\partial R_f^2} \bigg\lvert_{R_f = 0} R_f^2 + \mathcal{O}(R_f^3)= \bar{i}  + i R_f \frac{\partial \bar{i}}{\partial \eta} + R_f^2 \frac{1}{2}\left(\left(\frac{\partial \bar{i}}{\partial \eta}  \right)^2  + i\frac{\partial^2 \bar{i}}{\partial \eta^2}\right)+  \mathcal{O}(R_f^3).
\end{equation}
The ratio between the degraded and nondegraded currents in the second order is given by
\begin{equation}
    \frac{i}{\bar{i}} = \left(1- R_f\frac{\partial\bar{i}}{\partial \eta} - R_f^2 \frac{1}{2}\left(\bar{i}^{-1}\left(\frac{\partial \bar{i}}{\partial \eta}  \right)^2  + \frac{\partial^2 \bar{i}}{\partial \eta^2}\right)\right)^{-1} + \mathcal{O}(R_f^3). 
\end{equation}
The fitness value of the landscape is given by
\begin{equation}
     W(t,c,r, R_f)  \approx \frac{1}{r} \frac{1}{1- R_f \frac{\partial \bar{i}}{\partial \eta}-R_f^2 \frac{1}{2}\left(\bar{i}^{-1}\left(\frac{\partial \bar{i}}{\partial \eta}  \right)^2  + \frac{\partial^2 \bar{i}}{\partial \eta^2}\right)}.
\end{equation}
The second order model is not used in the simulations, but is given as an example to show how higher order terms would be derived.

\section{Analytical \change{Differential Conductance}}\label{ref:analyt_diff_res}

The transition state prefactor for the thermodynamically consistent Butler-Volmer equation is
\begin{equation}
    k_0(c) = k_0^* c^{\alpha} (a_+(1-c))^{1-\alpha},
\end{equation}
where $\alpha$ is the charge transfer coefficient, $a_+$ is the electrolyte activity coefficient, and $k_0^*$ is the current rate prefactor.
The analytical differential conductances are reported in the following sections where all values are reported in non-dimensionalized form.
In these functions, the exchange current density is actually the fitted prefactor \cite{fraggedakis2020scaling} that includes the lumped reaction rate prefactors.
For coupled ion-electron transfer, the formal overpotential is $e\eta_f = e\eta + k_B T\ln\frac{a_+}{c} $ to satisfy the De Donder relation \cite{Zhang2022_CIET_preprint}.
In future equations, it is assumed that the overpotential $\eta$ is nondimensionalized with the dimensional group $\frac{e}{k_B T}$.
In this series of equations, the helper function is found to be 
\begin{equation}
    \text{helper}(\eta_f, \lambda) = \frac{\sqrt{\lambda \pi}}{1+\exp{(-\eta_f)}} \text{erfc}\left(\frac{\lambda - \sqrt{1+\sqrt{\lambda} + \eta_f^2}}{2\sqrt{\lambda}} \right)
\end{equation}
and the derivative of the helper function with respect to the formal overpotential is found to be
\begin{equation}
    \frac{d\text{helper}}{d\eta_f}(\eta_f, \lambda) = \frac{\sqrt{\pi } \exp{(-\eta_f)}
   \sqrt{\lambda }
   \text{erfc}\left(\frac{\lambda
   -\sqrt{\eta_f^2+\sqrt{\lambda
   }+1}}{2 \sqrt{\lambda
   }}\right)+\frac{\left(\exp{(-\eta_f)}+1
   \right) \eta_f
   \exp{\left(-\frac{\left(\sqrt{\eta_f^2+\sqrt{\lambda }+1}-\lambda \right)^2}{4
   \lambda
   }\right)}}{\sqrt{\eta_f^2+\sqrt{\lambda
   }+1}}}{\left(\exp{(-\eta_f)}+1\right)^2
   }.
\end{equation}
In fact, though $\frac{d\text{helper}}{d\eta_f}(\eta_f, \lambda)$ is not strictly equal to $\frac{d\text{helper}}{d\eta}(\eta_f, \lambda)$, to the first order approximation of a Taylor expansion, if the perturbation to the electrolyte concentration and solid concentration at any given time is small, this solution is correct.
The derivation is shown in the following equation
\begin{equation}
     \frac{d\text{helper}}{d\eta} =  \frac{d\text{helper}}{d\eta_f- k_B Td\ln{a_+} + k_B T d\ln{c}} = \frac{d\text{helper}}{d\eta_f}\left[ 1+k_B T\frac{d\ln{a_+}}{d\eta_f} - k_B T\frac{d\ln{c}}{d\eta_f}\right] + \mathcal{O}(\Delta a_{(+)}^2) \approx \frac{d\text{helper}}{d\eta_f}.
\end{equation}
We thus use this approximation for most of our solutions, since perturbations to the electrolyte and solid lithium concentrations are not that large.

The second derivative of the helper function can also be found analytically to be
\begin{equation}
\begin{split}
\frac{d^2\text{helper}}{d\eta_f^2}(\eta_f, \lambda) & = 
    -\frac{1}{2
   \left(e^{-\eta_f}+1\right)^3}\left(-4 \sqrt{\pi } e^{-2\eta_{f}}
   \sqrt{\lambda}
   \text{erf}\left(\frac{\lambda-\sqrt{\eta_f^2
   +\sqrt{\lambda}+1}}{2
   \sqrt{\lambda}}\right)
   \right. \\
  & +  \left. 2 \sqrt{\pi } e^{-2
  \eta_f}
  \left(e^{\eta_f}+1\right) \sqrt{\lambda}
  \text{erf}\left(\frac{\lambda-\sqrt{\eta_f^2
  +\sqrt{\lambda}+1}}{2
  \sqrt{\lambda}}\right) \right. \\
  & +\frac{\eta_f^2
  \left(e^{-\eta_f}+1\right)^2
  e^{-\frac{\left(\sqrt{\eta_f^2+\sqrt{\lambda
  }+1}-\lambda\right)^2}
  {4 \lambda}}
  \left(\lambda-\sqrt{\eta_f^2+\sqrt{\lambda}+1}\right)}{\lambda \left(\eta_f^2+\sqrt{\lambda}+1\right)} \\
  & + \frac{2
  \left(e^{-\eta_f}+1\right)^2
  e^{-\frac{\left(\sqrt{\eta_f^2+\sqrt{\lambda
  }+1}-\lambda\right)^2}
  {4
  \lambda}}}{\sqrt{\eta_f^2+\sqrt{\lambda
  }+1}}-\frac{2 \eta_f^2
  \left(e^{-\eta_f}+1\right)^2
  e^{-\frac{\left(\sqrt{\eta_f^2+\sqrt{\lambda
  }+1}-\lambda\right)^2}
  {4
  \lambda}}}{\left(\eta_f^2+\sqrt{\lambda
  }+1\right)^{3/2}} \\
  & \left.+\frac{4
  \left(e^{\eta_f}+1\right) \eta_f
  e^{-\frac{\left(\sqrt{\eta_f^2+\sqrt{\lambda
  }+1}-\lambda\right)^2}
  {4 \lambda}-2
  \eta_f}}{\sqrt{\eta_f^2+\sqrt{\lambda}+
  1}}\right).
   \end{split}
\end{equation}

\begin{table} 
\centering 
\renewcommand{\arraystretch}{2}
\begin{tabular}{|c| c| c|}
\hline 
Model & Reaction Model $i(c)$ & $\begin{aligned}\frac{\partial \bar{i}}{\partial \eta}\end{aligned}$ \\
\hline 
Butler-Volmer & 
$\begin{aligned}k_0 (c)\left( \exp{\left(-\alpha \eta\right)} - \exp{\left((1-\alpha)\eta\right)}\right)\end{aligned}$ & 
$ -k_0 (c) \left( \alpha \exp{(-\alpha \eta)} + (1-\alpha) \exp{\left( (1-\alpha)\eta\right)} \right) $ \\
\hline 
Coupled Ion Electron Transfer & $ \begin{aligned} k_0^* (1-c)\left(a_+\text{helper}(-\eta_f, \lambda) - c\text{helper}(\eta_f, \lambda)\right)\end{aligned}$ &
$ -k_0^* (1-c)\left(a_+\frac{d\text{helper}}{d\eta_f}(-\eta_f, \lambda) + c\frac{d\text{helper}}{d\eta_f}(\eta_f, \lambda)\right) $
\\ \hline 
\end{tabular}
\end{table}
\begin{table} 
\centering 
\renewcommand{\arraystretch}{2}
\begin{tabular}{|c| c|}
\hline 
Model & $\begin{aligned}\frac{\partial^2 \bar{i}}{\partial \eta^2}\end{aligned}$ \\ \hline
Butler-Volmer & $k_0(c) \left( \alpha^2 \exp{\left(-\alpha \eta\right)} - (1-\alpha)^2\exp{\left((1-\alpha)\eta\right)}\right)$ \\
\hline 
Coupled Ion Electron Transfer  &
$ k_0^* (1-c)\left(a_+\frac{d^2\text{helper}}{d\eta_f^2}(-\eta_f, \lambda) - c\frac{d^2\text{helper}}{d\eta_f^2}(\eta_f, \lambda)\right) $ \\
\hline 
\end{tabular}
\end{table}

\section{Partial Derivatives}
\label{ref:partial_derivs}
\subsection{Film Resistance}
For the film resistance model, the partial derivative with respect to the film resistance can be found as
\begin{equation}
   \label{eq:derivative} \frac{\partial i}{\partial R_{f}} = \frac{\partial i}{\partial \eta} \frac{\partial \eta}{\partial R_{f}} = i \frac{\partial \bar{i}}{\partial \eta},
\end{equation}
which naturally evokes the value of the \change{differential conductance}.
The Taylor expansion of the system with respect to $R_f$ is 
\begin{equation}
    i = \bar{i} +\frac{\partial i}{\partial R_f} \bigg\lvert_{R_f = 0} R_f + \mathcal{O}(R_f^2)= \bar{i}  + i R_f \frac{\partial \bar{i}}{\partial \eta} +  \mathcal{O}(R_f^2),
\end{equation}
where all derivatives are evaluated at no degradation (we will neglect the evaluation terms for some future derivatives).
By combining terms and dividing by $\bar{i}$, we see that
\begin{equation}
    \frac{i}{\bar{i}} = \frac{1}{1- R_f\frac{\partial\bar{i}}{\partial \eta} } + \mathcal{O}(R_f^2), 
\end{equation}
which gives
\begin{equation}
    W  = \frac{1}{r} \left(1-R_f \frac{\partial \bar{i}}{\partial \eta}  \right)^{-1} + \mathcal{O}(R_f^2).
\end{equation}

For a symmetric Butler-Volmer model, it becomes more convenient to separate the prefactor effect and the driving force effect as $i = k_0(c) h(c,\eta)$.
Thus, we can expand the system as
\begin{equation}
    \frac{i}{\bar{i}} = \frac{k_0}{\bar{k}_0} \frac{h}{\bar{h}} = \frac{h}{\bar{h}}.
\end{equation}
When linearizing a symmetric Butler-Volmer model, a Taylor expansion of the driving force term shows that
\begin{equation}
\begin{split}
    h(\eta,i, R_f) 
    = \bar{h}(\eta)\left[\left( \sum_{k = 2n, n \in \mathbb{N}} \frac{(\alpha i R_f)^k}{k!}\right) + \left( \sum_{k = 2n+1, n \in \mathbb{N}} \frac{ (\alpha iR_f)^k}{k!}\right)\coth{\alpha \eta} \right].
    \label{eq:sym_BV_expand}
\end{split}
\end{equation}
We see that to the first order,
\begin{equation}
\begin{split}
   W  = \frac{1}{r}\left[1 + \alpha i \coth{(\alpha \eta)} R_f\right] + \mathcal{O}(R_f^2)
\end{split}
\end{equation}

\subsection{Surface Blockage}

For the surface blockage model, we separate the reaction rate effects into the prefactor effects from $k_0(c)$ and the reaction effects from $h(\eta)$.
Similarly to the approximation for resistance formation (but using $h$ instead of $i$), a Taylor expansion can be performed on $h(\eta)$ with respect to $\tilde{c}-1$ so that
\begin{equation}
\label{eq:h_surf_block_TS}
    h = \bar{h} + \frac{\partial h}{\partial \tilde{c}} \bigg|_{\tilde{c} = 1} (\tilde{c}-1) + \mathcal{O}(\tilde{c}^2) \approx \bar{h} + \frac{c(1-\tilde{c})}{(\tilde{c}-1)^2} \frac{\partial \mu_c}{\partial c} \frac{\partial \bar{h}}{\partial \eta}.
\end{equation}
Again, we see a form of \change{differential conductance} appear in this solution, since the \change{differential conductance} without the transition state prefactor can be obtained with the following relation
\begin{equation}
    \frac{\partial h}{\partial \tilde{c}} = \frac{\partial h}{\partial \eta} \frac{\partial \eta}{\partial \tilde{c}} = - \frac{c}{\tilde{c}^2} \frac{\partial \mu_c}{\partial c} \frac{\partial \bar{h}}{\partial \eta}.
\end{equation}
If we divide both sides of Eq. \ref{eq:h_surf_block_TS} by $\bar{h}$, we see that the ratio of the degraded to nondegraded reaction rate without the prefactor is found to be 
\begin{equation}
    \frac{h}{\bar{h}} \approx 1 + \frac{1}{\bar{h}} \frac{c(1-\tilde{c})}{\tilde{c}^2} \frac{\partial \mu_c}{\partial c} \frac{\partial \bar{h}}{\partial \eta}.
\end{equation}
In addition, the effects from $k_0(c)$ are shown as
\begin{equation}
    \frac{k_0(c)}{\bar{k}_{0}(c)} = \frac{\tilde{c}-c}{1-c}
\end{equation} 
for CIET and 
\begin{equation}
    \frac{k_0(c)}{\bar{k}_{0}(c)} = \left(\frac{\tilde{c}-c}{1-c}\right)^{1-\alpha}
\end{equation}
for the BV reaction.
Combining the two effects, we see that the fitness for the surface blockage model is
\begin{equation}
    W \approx \frac{1}{r} \left(\frac{\tilde{c}-c}{1-c} \right)^n \left(1 + \frac{1}{\bar{h}} \frac{c(1-\tilde{c})}{\tilde{c}^2} \frac{\partial \mu_c}{\partial c} \frac{\partial \bar{h}}{\partial \eta} \right),
\end{equation}
where $n = 1-\alpha$ for BV and $n = 1$ for CIET.
The effects of the latter term can be neglected if we focus on surface effects, so the approximation is
\begin{equation}
    W = \frac{1}{r} \left(\frac{\tilde{c}-c}{1-c} \right)^n + \mathcal{O}\left(\left(1-\tilde{c}\right)^2\right).
\end{equation}

\subsection{Electrolyte Loss}
For the \change{electrolyte loss} model, we operate exactly as we did in the surface blockage model, separating the reaction into the prefactor and the driving force components $i = k_0 h$ for a Butler-Volmer system.
We can separate the contributions from the transition state (performing a Taylor expansion on $c_+-1$) and the non-transition state effects
\begin{equation}
    h = \bar{h} + \frac{\partial h}{\partial c_+} \bigg|_{c_+ = 1}(c_+-1) + \mathcal{O}(\tilde{c}^2) \approx \bar{h} + \frac{\partial h}{\partial \eta}  \bigg|_{c_+ = 1}\frac{\partial \eta}{\partial c_+}\bigg|_{c_+ = 1}(c_+-1) = \bar{h} + \frac{k_B T}{e}\frac{\partial \ln{a_+}}{\partial \ln{c_+}}\frac{\partial \bar{h}}{\partial \eta} (1-c_+).
\end{equation}
For the overpotential for BV, $\frac{\partial \eta}{\partial c_+} = -\frac{k_B T}{ec_+}\frac{\partial \ln{a_+}}{\partial \ln{c_+}}$,
while for a coupled-ion electron transfer kinetic system, there is no dependence on formal overpotential on the electrolyte concentration ($\frac{\partial \eta_f}{\partial c_+} = 0$).
We also define the thermodynamic factor as $\frac{\partial \ln{a_+}}{\partial \ln{c_+}}$, which will be used in concentrated solution models \cite{newman2012electrochemical}.
We see that
\begin{equation}
    \frac{h}{\bar{h}} = \left[1 +\frac{1}{\bar{h}} \frac{\partial \ln{a_+}}{\partial \ln{c_+}} \frac{k_B T}{e}\frac{\partial \bar{h}}{\partial \eta} (1-c_+) \right],
\end{equation}
and since $\frac{k_0}{\bar{k}_0} = a_+^{1-\alpha}$, easily
\begin{equation}
    W = \frac{a_+^{1-\alpha}}{r}\left[1 +\frac{1}{\bar{h}}\frac{\partial \ln{a_+}}{\partial \ln{c_+}} \frac{k_B T}{e}\frac{\partial \bar{h}}{\partial \eta} (1-c_+) \right].
\end{equation}

For a coupled-ion electron transfer reaction, a Taylor expansion needs to be performed directly on the electrolyte concentration on $i$ since the effect of the electrolyte is convoluted throughout the reaction rate.
We obtain
\begin{equation}
    i = \bar{i} + \frac{di}{d c_+}(c_+-1) + \mathcal{O}\left((1-c_+)^2\right)
\end{equation}
where the full derivative of the current with respect to electrolyte concentration is  
\begin{equation}
    \frac{d i}{d c_+} = \frac{\partial i}{\partial \eta_f} \frac{\partial \eta_f}{\partial c_+} + \frac{\partial i}{\partial c_+},
\end{equation}
with the \change{differential conductance} seen again.
Since there is no dependence of the electrolyte concentration on the formal overpotential approximation ($\frac{\partial \eta_f}{\partial c_+} = 0 $), the electrolyte concentration dependence purely affects the reduction reaction in the coupled-ion electron transfer formulation $\frac{\partial i}{\partial c_+} = \frac{i_{red}} {c_+} \frac{\partial \ln{a_+}}{\partial \ln{c_+}}$, where $i_{red} = k_0^* (1-c) a_+\text{helper}(-\eta_f ,\lambda)$ is the reduction contribution to the driving force.
For a Stefan-Maxwell formulation, we see that
\begin{equation}
    \frac{di}{dc_+} =i_{red}c_+^{-1}\frac{\partial \ln{a_+}}{\partial \ln{c_+}},
\end{equation}
and thus, we see that
\begin{equation}
    \frac{i}{\bar{i}} = \left[1- \frac{\bar{i}_{red}}{\bar{i}}\left(1-c_+\right)\frac{\partial \ln{a_+}}{\partial \ln{c_+}}\right],
\end{equation}
which gives
\begin{equation}
    W = \frac{1}{r}\left[1- \frac{\bar{i}_{red}}{\bar{i}}\left(1-c_+\right) \frac{\partial \ln{a_+}}{\partial \ln{c_+}}\right]  + \mathcal{O}\left(((1-c_+)^2\right).
\end{equation}

\section{Simulation Parameters}\label{ref:appdx_sim_params}
An open circuit voltage model from Ref. \cite{colclasure2020electrode} was used for the NMC532 solid active material.
A constant exchange current density of $k_0^* = \SI{10}{A/m^2}$ for the intercalation reaction, while the reorganization energy in MHC/Marcus kinetics was found to be \SI{3.78}{k_BT} \cite{Zhang2022_CIET_preprint}.
The thermal diffusivity parameter in the system $D_0$ was set to be \SI{0.05}{k_B T}, and the degradation reaction formation voltage was set to \SI{4.1}{V} for all degradation mechanisms considered \cite{li2020degradation}.
The generalized Butler-Volmer reaction rate \cite{bazant2013theory} or coupled ion electron transfer reaction \cite{fraggedakis2021theory} were used to model the electrochemical ion insertion reaction.
A current control system with a C-rate of \SI{1}{C} was used in each of the model systems starting at a concentration of $0.45$, charging from $\SI{3.7}{V}$ to $\SI{4.1}{V}$ until the battery dies.
Discretizations of $0.002$ for the concentration and \SI{10}{nm} were used for the radius, which was discretized from \SI{10}{nm} to \SI{300}{nm}.
A normal distribution with an average of \SI{100}{nm} and a variance of \SI{100}{nm} was selected to perform this set of simulations, which is an abnormally large distribution used so that a wide variance of particle sizes could be sampled.
For electrolyte, a dilute solution model of $1$ M is used.
The BV reaction rate as well as the localized and delocalized electron limits of CIET were used to study this system, which are displayed in Appendix A.
The fitness values were used with the approximation from Eq. \ref{eq:approximation1}.
The scripts used to run this set of simulations can be found in the public repository \textit{lightningclaw001/public\_paper\_scripts} under the folder \textit{fitness\_distribution} (\url{https://github.com/lightningclaw001/public_paper_scripts/tree/main/fitness_distribution}) for the different models.

The degradation parameters are listed below. 
A resistivity of  $\SI{1}{\Omega \cdot m^4/C}$ was used in the resistive film simulations for the film material.
The exchange current density of the degradation reaction for the BV reaction for resistive film formation was set to \SI{0.04}{A/m^2}, while for the CIET reaction it was set to \SI{0.2}{A/m^2}.
The exchange current density of the degradation reaction $k_{0,deg}$ was set to $0.03$\SI{}{A/m^2} for the surface blockage mechanism reactions.
For the \change{electrolyte loss} reaction mechanism, the coefficient for \change{electrolyte loss} $k$ was set to \SI{0.003}{M/hr}.
The capacity loss and \change{electrolyte loss} models were chosen so that 90\% of the original capacity is achieved roughly at the end of lifetime.

For all degradation models, the total intercalation current is then found to be
\begin{equation}
     \langle \text{Rxn}_{int} \rangle = \frac{d \langle c \rangle}{dt} = \frac{\int_0^\infty \int_0^1 c r^3 \frac{df}{dt} dcdr}{\int_0^\infty r^3 g(r) dr}
    \end{equation}
integrated over the total volume of the system, where $g(r) = \int_0^1 f(c,r)dc$ is the constant probability distribution of the particle sizes.
The total degradation current is then similarly found to be 
\begin{equation}
     \langle \text{Rxn}_{deg} \rangle  = \frac{\int_0^\infty \int_0^1 c r^2 i_{deg} dcdr}{ \int_0^\infty r^3 g(r) dr} = \frac{\int_0^\infty r^2 \langle i_{deg}\rangle g(r)dr}{ \int_0^\infty r^3 g(r) dr},
    \end{equation}
since in the system, we only calculate the averaged amount of degradation over all particles of the same size $g(r)$.
Thus, the total applied current in the system is system is found to be 
$\langle \text{Rxn}_{constraint} \rangle = \langle \text{Rxn}_{int} \rangle + \langle \text{Rxn}_{deg} \rangle $.

Because the Butler-Volmer reaction grows exponentially as the overpotential in the system increases, the large values of the overpotential cause artificial numerical errors to be introduced because of the magnitude of reaction rates at high overpotentials.
Thus, it becomes necessary to add a damping function multiplied to the reaction rates at the high overpotential terms to prevent this from happening, causing the Fokker-Planck equation to become 
\begin{equation}
    \frac{\partial f}{\partial t} = - \frac{\partial }{\partial c}\left(fiW\zeta \right) + \frac{\partial }{\partial c} \left( D\frac{\partial }{\partial c}\left( fW \zeta\right)\right)
\end{equation}
Because the material is not thermodynamically phase separating, there is a very low density of the population at these high overpotential concentrations, so the damping function does not affect the solution of the system.
We choose a damping function
\begin{equation}
    \zeta(\eta,z) = \frac{1}{2}\left(\tanh{\left(-(|\eta|-z)\right)}+1\right),
\end{equation}
where $z$ is the cutoff value for the overpotential, which we set to $z = \SI{8}{k_BT}$.
This damping function is symmetric with respect to $\eta = 0$ and dampens the overlarge values of reaction rate caused by the unphysically high overpotential from Butler-Volmer.

Because of the difficulty of numerically solving the reaction rate for many implicit solutions as would be required especially for a model with the resistive film buildup, we instead turn to our analytical approximations performed in Sec. \ref{sec:resistive_film} and use the $\mathcal{O}(3)$ approximation for the Butler-Volmer equation in Eq. \ref{eq:fitness_value1}, and the  $\mathcal{O}(2)$ approximation for the coupled ion electron transfer solution in Eq. \ref{eq:approximation1} for the Fokker-Planck Eq. \ref{eq:approx_FP}.
The errors to the numerical approximations are shown in Appendix \ref{ref:appdx_numerical_error}.
The other degradation mechanisms of surface blockage and \change{electrolyte loss} have simpler equations to solve and do not need implicit solutions for the intercalation rates, so their full solutions to Eq. \ref{eq:full_FP} are used in the simulations.
In the full model with all three degradation models, since it is necessary to solve the implicit reaction rate, we apply the approximations to the fitness function in Eqs. \ref{eq:W_BV} and \ref{eq:W_CIET}.

\section{Numerical Error} \label{ref:appdx_numerical_error}
The $\mathcal{O}(4)$ error for the Butler-Volmer resistance reaction or the $\mathcal{O}(2)$ error for the other reaction rates and degradation mechanisms are shown in Fig. \ref{fig:appdx_fig1} for the approximate solutions to the analytical solutions.
\begin{figure*}[t]
\includegraphics[width=0.5\textwidth]{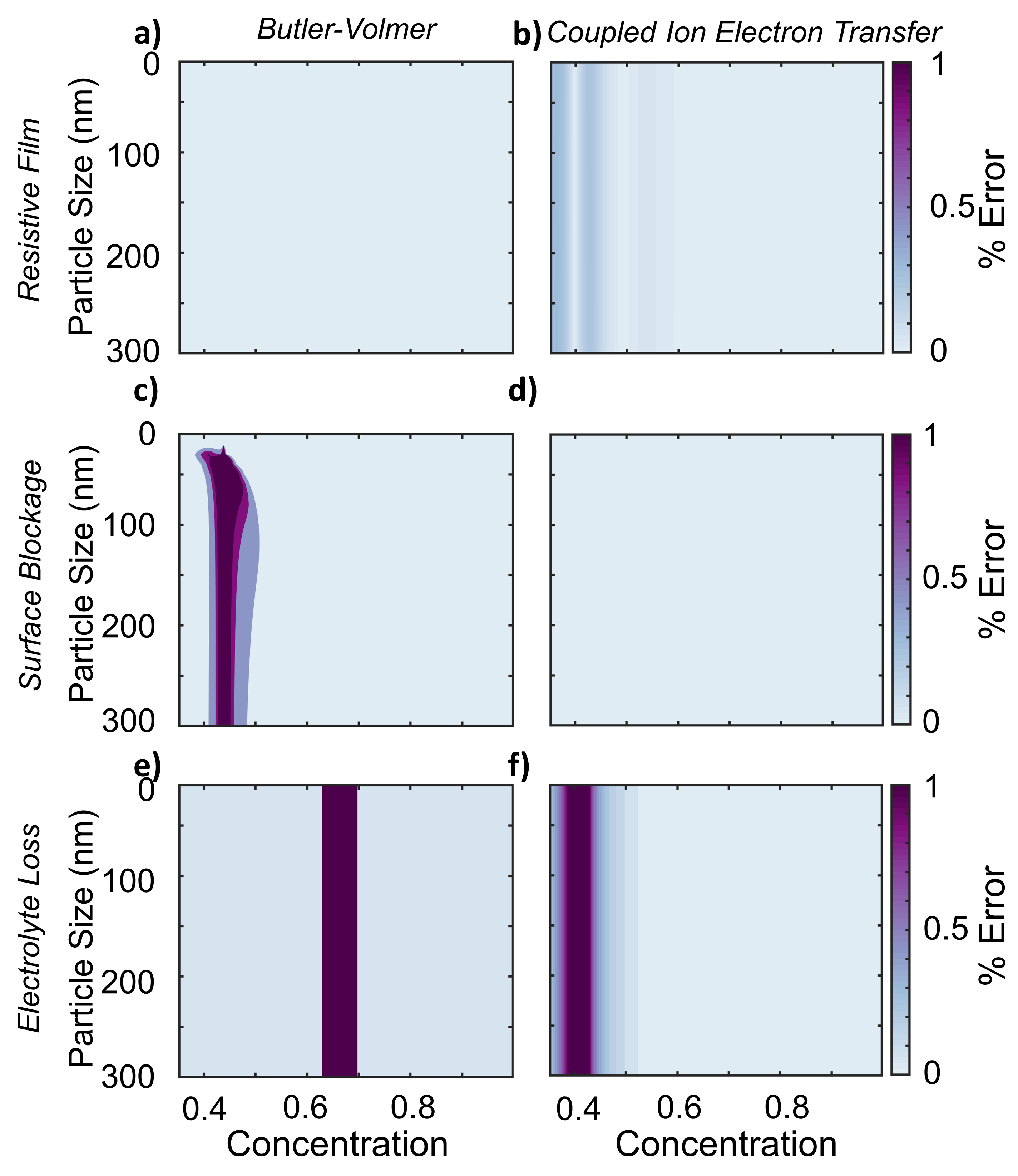}
\centering
\caption{The percentage errors of the analytical solutions are plotted below for each reaction mechanism and degradation mechanism.}
\label{fig:appdx_fig1}
\end{figure*}
The exact degradation mechanisms are simulated for the surface blockage and the \change{electrolyte loss} solutions, but the error of the analytical solutions are provided for reference.

 \end{widetext}

\bibliography{apssamp}

\begin{thebibliography}{81}%
\makeatletter
\providecommand \@ifxundefined [1]{%
 \@ifx{#1\undefined}
}%
\providecommand \@ifnum [1]{%
 \ifnum #1\expandafter \@firstoftwo
 \else \expandafter \@secondoftwo
 \fi
}%
\providecommand \@ifx [1]{%
 \ifx #1\expandafter \@firstoftwo
 \else \expandafter \@secondoftwo
 \fi
}%
\providecommand \natexlab [1]{#1}%
\providecommand \enquote  [1]{``#1''}%
\providecommand \bibnamefont  [1]{#1}%
\providecommand \bibfnamefont [1]{#1}%
\providecommand \citenamefont [1]{#1}%
\providecommand \href@noop [0]{\@secondoftwo}%
\providecommand \href [0]{\begingroup \@sanitize@url \@href}%
\providecommand \@href[1]{\@@startlink{#1}\@@href}%
\providecommand \@@href[1]{\endgroup#1\@@endlink}%
\providecommand \@sanitize@url [0]{\catcode `\\12\catcode `\$12\catcode
  `\&12\catcode `\#12\catcode `\^12\catcode `\_12\catcode `\%12\relax}%
\providecommand \@@startlink[1]{}%
\providecommand \@@endlink[0]{}%
\providecommand \url  [0]{\begingroup\@sanitize@url \@url }%
\providecommand \@url [1]{\endgroup\@href {#1}{\urlprefix }}%
\providecommand \urlprefix  [0]{URL }%
\providecommand \Eprint [0]{\href }%
\providecommand \doibase [0]{https://doi.org/}%
\providecommand \selectlanguage [0]{\@gobble}%
\providecommand \bibinfo  [0]{\@secondoftwo}%
\providecommand \bibfield  [0]{\@secondoftwo}%
\providecommand \translation [1]{[#1]}%
\providecommand \BibitemOpen [0]{}%
\providecommand \bibitemStop [0]{}%
\providecommand \bibitemNoStop [0]{.\EOS\space}%
\providecommand \EOS [0]{\spacefactor3000\relax}%
\providecommand \BibitemShut  [1]{\csname bibitem#1\endcsname}%
\let\auto@bib@innerbib\@empty
\bibitem [{\citenamefont {Li}\ \emph {et~al.}(2018{\natexlab{a}})\citenamefont
  {Li}, \citenamefont {Lu}, \citenamefont {Chen},\ and\ \citenamefont
  {Amine}}]{li201830}%
  \BibitemOpen
  \bibfield  {author} {\bibinfo {author} {\bibfnamefont {M.}~\bibnamefont
  {Li}}, \bibinfo {author} {\bibfnamefont {J.}~\bibnamefont {Lu}}, \bibinfo
  {author} {\bibfnamefont {Z.}~\bibnamefont {Chen}},\ and\ \bibinfo {author}
  {\bibfnamefont {K.}~\bibnamefont {Amine}},\ }\bibfield  {title} {\bibinfo
  {title} {30 years of lithium-ion batteries},\ }\href@noop {} {\bibfield
  {journal} {\bibinfo  {journal} {Advanced Materials}\ }\textbf {\bibinfo
  {volume} {30}},\ \bibinfo {pages} {1800561} (\bibinfo {year}
  {2018}{\natexlab{a}})}\BibitemShut {NoStop}%
\bibitem [{\citenamefont {Kabir}\ and\ \citenamefont
  {Demirocak}(2017)}]{kabir2017degradation}%
  \BibitemOpen
  \bibfield  {author} {\bibinfo {author} {\bibfnamefont {M.}~\bibnamefont
  {Kabir}}\ and\ \bibinfo {author} {\bibfnamefont {D.~E.}\ \bibnamefont
  {Demirocak}},\ }\bibfield  {title} {\bibinfo {title} {Degradation mechanisms
  in li-ion batteries: a state-of-the-art review},\ }\href@noop {} {\bibfield
  {journal} {\bibinfo  {journal} {International Journal of Energy Research}\
  }\textbf {\bibinfo {volume} {41}},\ \bibinfo {pages} {1963} (\bibinfo {year}
  {2017})}\BibitemShut {NoStop}%
\bibitem [{\citenamefont {Goodenough}\ and\ \citenamefont
  {Park}(2013)}]{goodenough2013li}%
  \BibitemOpen
  \bibfield  {author} {\bibinfo {author} {\bibfnamefont {J.~B.}\ \bibnamefont
  {Goodenough}}\ and\ \bibinfo {author} {\bibfnamefont {K.-S.}\ \bibnamefont
  {Park}},\ }\bibfield  {title} {\bibinfo {title} {The li-ion rechargeable
  battery: a perspective},\ }\href@noop {} {\bibfield  {journal} {\bibinfo
  {journal} {Journal of the American Chemical Society}\ }\textbf {\bibinfo
  {volume} {135}},\ \bibinfo {pages} {1167} (\bibinfo {year}
  {2013})}\BibitemShut {NoStop}%
\bibitem [{\citenamefont {Berliner}\ \emph {et~al.}(2021)\citenamefont
  {Berliner}, \citenamefont {Zhao}, \citenamefont {Das}, \citenamefont
  {Forsuelo}, \citenamefont {Jiang}, \citenamefont {Chueh}, \citenamefont
  {Bazant},\ and\ \citenamefont {Braatz}}]{berliner2021nonlinear}%
  \BibitemOpen
  \bibfield  {author} {\bibinfo {author} {\bibfnamefont {M.~D.}\ \bibnamefont
  {Berliner}}, \bibinfo {author} {\bibfnamefont {H.}~\bibnamefont {Zhao}},
  \bibinfo {author} {\bibfnamefont {S.}~\bibnamefont {Das}}, \bibinfo {author}
  {\bibfnamefont {M.}~\bibnamefont {Forsuelo}}, \bibinfo {author}
  {\bibfnamefont {B.}~\bibnamefont {Jiang}}, \bibinfo {author} {\bibfnamefont
  {W.~H.}\ \bibnamefont {Chueh}}, \bibinfo {author} {\bibfnamefont {M.~Z.}\
  \bibnamefont {Bazant}},\ and\ \bibinfo {author} {\bibfnamefont {R.~D.}\
  \bibnamefont {Braatz}},\ }\bibfield  {title} {\bibinfo {title} {Nonlinear
  identifiability analysis of the porous electrode theory model of lithium-ion
  batteries},\ }\href@noop {} {\bibfield  {journal} {\bibinfo  {journal}
  {Journal of The Electrochemical Society}\ }\textbf {\bibinfo {volume}
  {168}},\ \bibinfo {pages} {090546} (\bibinfo {year} {2021})}\BibitemShut
  {NoStop}%
\bibitem [{\citenamefont {Attia}\ \emph {et~al.}(2019)\citenamefont {Attia},
  \citenamefont {Das}, \citenamefont {Harris}, \citenamefont {Bazant},\ and\
  \citenamefont {Chueh}}]{attia2019electrochemical}%
  \BibitemOpen
  \bibfield  {author} {\bibinfo {author} {\bibfnamefont {P.~M.}\ \bibnamefont
  {Attia}}, \bibinfo {author} {\bibfnamefont {S.}~\bibnamefont {Das}}, \bibinfo
  {author} {\bibfnamefont {S.~J.}\ \bibnamefont {Harris}}, \bibinfo {author}
  {\bibfnamefont {M.~Z.}\ \bibnamefont {Bazant}},\ and\ \bibinfo {author}
  {\bibfnamefont {W.~C.}\ \bibnamefont {Chueh}},\ }\bibfield  {title} {\bibinfo
  {title} {Electrochemical kinetics of sei growth on carbon black: Part i.
  experiments},\ }\href@noop {} {\bibfield  {journal} {\bibinfo  {journal}
  {Journal of the Electrochemical Society}\ }\textbf {\bibinfo {volume}
  {166}},\ \bibinfo {pages} {E97} (\bibinfo {year} {2019})}\BibitemShut
  {NoStop}%
\bibitem [{\citenamefont {Das}\ \emph {et~al.}(2019)\citenamefont {Das},
  \citenamefont {Attia}, \citenamefont {Chueh},\ and\ \citenamefont
  {Bazant}}]{das2019electrochemical}%
  \BibitemOpen
  \bibfield  {author} {\bibinfo {author} {\bibfnamefont {S.}~\bibnamefont
  {Das}}, \bibinfo {author} {\bibfnamefont {P.~M.}\ \bibnamefont {Attia}},
  \bibinfo {author} {\bibfnamefont {W.~C.}\ \bibnamefont {Chueh}},\ and\
  \bibinfo {author} {\bibfnamefont {M.~Z.}\ \bibnamefont {Bazant}},\ }\bibfield
   {title} {\bibinfo {title} {Electrochemical kinetics of sei growth on carbon
  black: Part ii. modeling},\ }\href@noop {} {\bibfield  {journal} {\bibinfo
  {journal} {Journal of The Electrochemical Society}\ }\textbf {\bibinfo
  {volume} {166}},\ \bibinfo {pages} {E107} (\bibinfo {year}
  {2019})}\BibitemShut {NoStop}%
\bibitem [{\citenamefont {Yan}\ \emph {et~al.}(2015)\citenamefont {Yan},
  \citenamefont {Nie}, \citenamefont {Zheng}, \citenamefont {Zhou},
  \citenamefont {Lu}, \citenamefont {Zhang}, \citenamefont {Xu}, \citenamefont
  {Belharouak}, \citenamefont {Zu}, \citenamefont {Xiao} \emph
  {et~al.}}]{yan2015evolution}%
  \BibitemOpen
  \bibfield  {author} {\bibinfo {author} {\bibfnamefont {P.}~\bibnamefont
  {Yan}}, \bibinfo {author} {\bibfnamefont {A.}~\bibnamefont {Nie}}, \bibinfo
  {author} {\bibfnamefont {J.}~\bibnamefont {Zheng}}, \bibinfo {author}
  {\bibfnamefont {Y.}~\bibnamefont {Zhou}}, \bibinfo {author} {\bibfnamefont
  {D.}~\bibnamefont {Lu}}, \bibinfo {author} {\bibfnamefont {X.}~\bibnamefont
  {Zhang}}, \bibinfo {author} {\bibfnamefont {R.}~\bibnamefont {Xu}}, \bibinfo
  {author} {\bibfnamefont {I.}~\bibnamefont {Belharouak}}, \bibinfo {author}
  {\bibfnamefont {X.}~\bibnamefont {Zu}}, \bibinfo {author} {\bibfnamefont
  {J.}~\bibnamefont {Xiao}}, \emph {et~al.},\ }\bibfield  {title} {\bibinfo
  {title} {Evolution of lattice structure and chemical composition of the
  surface reconstruction layer in li1. 2ni0. 2mn0. 6o2 cathode material for
  lithium ion batteries},\ }\href@noop {} {\bibfield  {journal} {\bibinfo
  {journal} {Nano letters}\ }\textbf {\bibinfo {volume} {15}},\ \bibinfo
  {pages} {514} (\bibinfo {year} {2015})}\BibitemShut {NoStop}%
\bibitem [{\citenamefont {Zheng}\ \emph {et~al.}(2019)\citenamefont {Zheng},
  \citenamefont {Ye}, \citenamefont {Liu}, \citenamefont {Xiao}, \citenamefont
  {Wang}, \citenamefont {Wang},\ and\ \citenamefont {Pan}}]{zheng2019ni}%
  \BibitemOpen
  \bibfield  {author} {\bibinfo {author} {\bibfnamefont {J.}~\bibnamefont
  {Zheng}}, \bibinfo {author} {\bibfnamefont {Y.}~\bibnamefont {Ye}}, \bibinfo
  {author} {\bibfnamefont {T.}~\bibnamefont {Liu}}, \bibinfo {author}
  {\bibfnamefont {Y.}~\bibnamefont {Xiao}}, \bibinfo {author} {\bibfnamefont
  {C.}~\bibnamefont {Wang}}, \bibinfo {author} {\bibfnamefont {F.}~\bibnamefont
  {Wang}},\ and\ \bibinfo {author} {\bibfnamefont {F.}~\bibnamefont {Pan}},\
  }\bibfield  {title} {\bibinfo {title} {Ni/li disordering in layered
  transition metal oxide: electrochemical impact, origin, and control},\
  }\href@noop {} {\bibfield  {journal} {\bibinfo  {journal} {Accounts of
  chemical research}\ }\textbf {\bibinfo {volume} {52}},\ \bibinfo {pages}
  {2201} (\bibinfo {year} {2019})}\BibitemShut {NoStop}%
\bibitem [{\citenamefont {Sarasketa-Zabala}\ \emph {et~al.}(2015)\citenamefont
  {Sarasketa-Zabala}, \citenamefont {Aguesse}, \citenamefont {Villarreal},
  \citenamefont {Rodriguez-Martinez}, \citenamefont {L{\'o}pez},\ and\
  \citenamefont {Kubiak}}]{sarasketa2015understanding}%
  \BibitemOpen
  \bibfield  {author} {\bibinfo {author} {\bibfnamefont {E.}~\bibnamefont
  {Sarasketa-Zabala}}, \bibinfo {author} {\bibfnamefont {F.}~\bibnamefont
  {Aguesse}}, \bibinfo {author} {\bibfnamefont {I.}~\bibnamefont {Villarreal}},
  \bibinfo {author} {\bibfnamefont {L.}~\bibnamefont {Rodriguez-Martinez}},
  \bibinfo {author} {\bibfnamefont {C.~M.}\ \bibnamefont {L{\'o}pez}},\ and\
  \bibinfo {author} {\bibfnamefont {P.}~\bibnamefont {Kubiak}},\ }\bibfield
  {title} {\bibinfo {title} {Understanding lithium inventory loss and sudden
  performance fade in cylindrical cells during cycling with deep-discharge
  steps},\ }\href@noop {} {\bibfield  {journal} {\bibinfo  {journal} {The
  Journal of Physical Chemistry C}\ }\textbf {\bibinfo {volume} {119}},\
  \bibinfo {pages} {896} (\bibinfo {year} {2015})}\BibitemShut {NoStop}%
\bibitem [{\citenamefont {Nanda}\ and\ \citenamefont
  {Manthiram}(2020)}]{nanda2020lithium}%
  \BibitemOpen
  \bibfield  {author} {\bibinfo {author} {\bibfnamefont {S.}~\bibnamefont
  {Nanda}}\ and\ \bibinfo {author} {\bibfnamefont {A.}~\bibnamefont
  {Manthiram}},\ }\bibfield  {title} {\bibinfo {title} {Lithium degradation in
  lithium--sulfur batteries: insights into inventory depletion and interphasial
  evolution with cycling},\ }\href@noop {} {\bibfield  {journal} {\bibinfo
  {journal} {Energy \& Environmental Science}\ }\textbf {\bibinfo {volume}
  {13}},\ \bibinfo {pages} {2501} (\bibinfo {year} {2020})}\BibitemShut
  {NoStop}%
\bibitem [{\citenamefont {Severson}\ \emph {et~al.}(2019)\citenamefont
  {Severson}, \citenamefont {Attia}, \citenamefont {Jin}, \citenamefont
  {Perkins}, \citenamefont {Jiang}, \citenamefont {Yang}, \citenamefont {Chen},
  \citenamefont {Aykol}, \citenamefont {Herring}, \citenamefont {Fraggedakis}
  \emph {et~al.}}]{severson2019data}%
  \BibitemOpen
  \bibfield  {author} {\bibinfo {author} {\bibfnamefont {K.~A.}\ \bibnamefont
  {Severson}}, \bibinfo {author} {\bibfnamefont {P.~M.}\ \bibnamefont {Attia}},
  \bibinfo {author} {\bibfnamefont {N.}~\bibnamefont {Jin}}, \bibinfo {author}
  {\bibfnamefont {N.}~\bibnamefont {Perkins}}, \bibinfo {author} {\bibfnamefont
  {B.}~\bibnamefont {Jiang}}, \bibinfo {author} {\bibfnamefont
  {Z.}~\bibnamefont {Yang}}, \bibinfo {author} {\bibfnamefont {M.~H.}\
  \bibnamefont {Chen}}, \bibinfo {author} {\bibfnamefont {M.}~\bibnamefont
  {Aykol}}, \bibinfo {author} {\bibfnamefont {P.~K.}\ \bibnamefont {Herring}},
  \bibinfo {author} {\bibfnamefont {D.}~\bibnamefont {Fraggedakis}}, \emph
  {et~al.},\ }\bibfield  {title} {\bibinfo {title} {Data-driven prediction of
  battery cycle life before capacity degradation},\ }\href@noop {} {\bibfield
  {journal} {\bibinfo  {journal} {Nature Energy}\ }\textbf {\bibinfo {volume}
  {4}},\ \bibinfo {pages} {383} (\bibinfo {year} {2019})}\BibitemShut {NoStop}%
\bibitem [{\citenamefont {Ma}\ \emph {et~al.}(2019)\citenamefont {Ma},
  \citenamefont {Harlow}, \citenamefont {Li}, \citenamefont {Ma}, \citenamefont
  {Hall}, \citenamefont {Buteau}, \citenamefont {Genovese}, \citenamefont
  {Cormier},\ and\ \citenamefont {Dahn}}]{ma2019hindering}%
  \BibitemOpen
  \bibfield  {author} {\bibinfo {author} {\bibfnamefont {X.}~\bibnamefont
  {Ma}}, \bibinfo {author} {\bibfnamefont {J.~E.}\ \bibnamefont {Harlow}},
  \bibinfo {author} {\bibfnamefont {J.}~\bibnamefont {Li}}, \bibinfo {author}
  {\bibfnamefont {L.}~\bibnamefont {Ma}}, \bibinfo {author} {\bibfnamefont
  {D.~S.}\ \bibnamefont {Hall}}, \bibinfo {author} {\bibfnamefont
  {S.}~\bibnamefont {Buteau}}, \bibinfo {author} {\bibfnamefont
  {M.}~\bibnamefont {Genovese}}, \bibinfo {author} {\bibfnamefont
  {M.}~\bibnamefont {Cormier}},\ and\ \bibinfo {author} {\bibfnamefont
  {J.}~\bibnamefont {Dahn}},\ }\bibfield  {title} {\bibinfo {title} {Hindering
  rollover failure of li [ni0. 5mn0. 3co0. 2] o2/graphite pouch cells during
  long-term cycling},\ }\href@noop {} {\bibfield  {journal} {\bibinfo
  {journal} {Journal of The Electrochemical Society}\ }\textbf {\bibinfo
  {volume} {166}},\ \bibinfo {pages} {A711} (\bibinfo {year}
  {2019})}\BibitemShut {NoStop}%
\bibitem [{\citenamefont {Attia}\ \emph {et~al.}(2022)\citenamefont {Attia},
  \citenamefont {Bills}, \citenamefont {Planella}, \citenamefont {Dechent},
  \citenamefont {Reis}, \citenamefont {Dubarry}, \citenamefont {Gasper},
  \citenamefont {Gilchrist}, \citenamefont {Greenbank}, \citenamefont {Howey}
  \emph {et~al.}}]{attia2022knees}%
  \BibitemOpen
  \bibfield  {author} {\bibinfo {author} {\bibfnamefont {P.~M.}\ \bibnamefont
  {Attia}}, \bibinfo {author} {\bibfnamefont {A.}~\bibnamefont {Bills}},
  \bibinfo {author} {\bibfnamefont {F.~B.}\ \bibnamefont {Planella}}, \bibinfo
  {author} {\bibfnamefont {P.}~\bibnamefont {Dechent}}, \bibinfo {author}
  {\bibfnamefont {G.~d.}\ \bibnamefont {Reis}}, \bibinfo {author}
  {\bibfnamefont {M.}~\bibnamefont {Dubarry}}, \bibinfo {author} {\bibfnamefont
  {P.}~\bibnamefont {Gasper}}, \bibinfo {author} {\bibfnamefont
  {R.}~\bibnamefont {Gilchrist}}, \bibinfo {author} {\bibfnamefont
  {S.}~\bibnamefont {Greenbank}}, \bibinfo {author} {\bibfnamefont
  {D.}~\bibnamefont {Howey}}, \emph {et~al.},\ }\bibfield  {title} {\bibinfo
  {title} {" knees" in lithium-ion battery aging trajectories},\ }\href@noop {}
  {\bibfield  {journal} {\bibinfo  {journal} {arXiv preprint arXiv:2201.02891}\
  } (\bibinfo {year} {2022})}\BibitemShut {NoStop}%
\bibitem [{\citenamefont
  {Nagasubramanian}(2000)}]{nagasubramanian2000impedance}%
  \BibitemOpen
  \bibfield  {author} {\bibinfo {author} {\bibfnamefont {G.}~\bibnamefont
  {Nagasubramanian}},\ }\href@noop {} {\emph {\bibinfo {title} {Impedance
  studies on Li-ion cathodes}}},\ \bibinfo {type} {Tech. Rep.}\ (\bibinfo
  {institution} {Sandia National Lab.(SNL-NM), Albuquerque, NM (United States);
  Sandia~…},\ \bibinfo {year} {2000})\BibitemShut {NoStop}%
\bibitem [{\citenamefont {Sallis}\ \emph {et~al.}(2016)\citenamefont {Sallis},
  \citenamefont {Pereira}, \citenamefont {Mukherjee}, \citenamefont
  {Quackenbush}, \citenamefont {Faenza}, \citenamefont {Schlueter},
  \citenamefont {Lee}, \citenamefont {Yang}, \citenamefont {Cosandey},
  \citenamefont {Amatucci} \emph {et~al.}}]{sallis2016surface}%
  \BibitemOpen
  \bibfield  {author} {\bibinfo {author} {\bibfnamefont {S.}~\bibnamefont
  {Sallis}}, \bibinfo {author} {\bibfnamefont {N.}~\bibnamefont {Pereira}},
  \bibinfo {author} {\bibfnamefont {P.}~\bibnamefont {Mukherjee}}, \bibinfo
  {author} {\bibfnamefont {N.}~\bibnamefont {Quackenbush}}, \bibinfo {author}
  {\bibfnamefont {N.}~\bibnamefont {Faenza}}, \bibinfo {author} {\bibfnamefont
  {C.}~\bibnamefont {Schlueter}}, \bibinfo {author} {\bibfnamefont {T.-L.}\
  \bibnamefont {Lee}}, \bibinfo {author} {\bibfnamefont {W.}~\bibnamefont
  {Yang}}, \bibinfo {author} {\bibfnamefont {F.}~\bibnamefont {Cosandey}},
  \bibinfo {author} {\bibfnamefont {G.}~\bibnamefont {Amatucci}}, \emph
  {et~al.},\ }\bibfield  {title} {\bibinfo {title} {Surface degradation of
  li1--x ni0. 80co0. 15al0. 05o2 cathodes: Correlating charge transfer
  impedance with surface phase transformations},\ }\href@noop {} {\bibfield
  {journal} {\bibinfo  {journal} {Applied Physics Letters}\ }\textbf {\bibinfo
  {volume} {108}},\ \bibinfo {pages} {263902} (\bibinfo {year}
  {2016})}\BibitemShut {NoStop}%
\bibitem [{\citenamefont {Gao}\ \emph {et~al.}(2021)\citenamefont {Gao},
  \citenamefont {Han}, \citenamefont {Fraggedakis}, \citenamefont {Das},
  \citenamefont {Zhou}, \citenamefont {Yeh}, \citenamefont {Xu}, \citenamefont
  {Chueh}, \citenamefont {Li},\ and\ \citenamefont
  {Bazant}}]{gao2021interplay}%
  \BibitemOpen
  \bibfield  {author} {\bibinfo {author} {\bibfnamefont {T.}~\bibnamefont
  {Gao}}, \bibinfo {author} {\bibfnamefont {Y.}~\bibnamefont {Han}}, \bibinfo
  {author} {\bibfnamefont {D.}~\bibnamefont {Fraggedakis}}, \bibinfo {author}
  {\bibfnamefont {S.}~\bibnamefont {Das}}, \bibinfo {author} {\bibfnamefont
  {T.}~\bibnamefont {Zhou}}, \bibinfo {author} {\bibfnamefont {C.-N.}\
  \bibnamefont {Yeh}}, \bibinfo {author} {\bibfnamefont {S.}~\bibnamefont
  {Xu}}, \bibinfo {author} {\bibfnamefont {W.~C.}\ \bibnamefont {Chueh}},
  \bibinfo {author} {\bibfnamefont {J.}~\bibnamefont {Li}},\ and\ \bibinfo
  {author} {\bibfnamefont {M.~Z.}\ \bibnamefont {Bazant}},\ }\bibfield  {title}
  {\bibinfo {title} {Interplay of lithium intercalation and plating on a single
  graphite particle},\ }\href@noop {} {\bibfield  {journal} {\bibinfo
  {journal} {Joule}\ }\textbf {\bibinfo {volume} {5}},\ \bibinfo {pages} {393}
  (\bibinfo {year} {2021})}\BibitemShut {NoStop}%
\bibitem [{\citenamefont {Mandli}\ \emph {et~al.}(2019)\citenamefont {Mandli},
  \citenamefont {Kaushik}, \citenamefont {Patil}, \citenamefont {Naha},
  \citenamefont {Hariharan}, \citenamefont {Kolake}, \citenamefont {Han},\ and\
  \citenamefont {Choi}}]{mandli2019analysis}%
  \BibitemOpen
  \bibfield  {author} {\bibinfo {author} {\bibfnamefont {A.~R.}\ \bibnamefont
  {Mandli}}, \bibinfo {author} {\bibfnamefont {A.}~\bibnamefont {Kaushik}},
  \bibinfo {author} {\bibfnamefont {R.~S.}\ \bibnamefont {Patil}}, \bibinfo
  {author} {\bibfnamefont {A.}~\bibnamefont {Naha}}, \bibinfo {author}
  {\bibfnamefont {K.~S.}\ \bibnamefont {Hariharan}}, \bibinfo {author}
  {\bibfnamefont {S.~M.}\ \bibnamefont {Kolake}}, \bibinfo {author}
  {\bibfnamefont {S.}~\bibnamefont {Han}},\ and\ \bibinfo {author}
  {\bibfnamefont {W.}~\bibnamefont {Choi}},\ }\bibfield  {title} {\bibinfo
  {title} {Analysis of the effect of resistance increase on the capacity fade
  of lithium ion batteries},\ }\href@noop {} {\bibfield  {journal} {\bibinfo
  {journal} {International Journal of Energy Research}\ }\textbf {\bibinfo
  {volume} {43}},\ \bibinfo {pages} {2044} (\bibinfo {year}
  {2019})}\BibitemShut {NoStop}%
\bibitem [{\citenamefont {Ott}\ and\ \citenamefont
  {Antonsen~Jr}(2017)}]{ott2017frequency}%
  \BibitemOpen
  \bibfield  {author} {\bibinfo {author} {\bibfnamefont {E.}~\bibnamefont
  {Ott}}\ and\ \bibinfo {author} {\bibfnamefont {T.~M.}\ \bibnamefont
  {Antonsen~Jr}},\ }\bibfield  {title} {\bibinfo {title} {Frequency and phase
  synchronization in large groups: Low dimensional description of synchronized
  clapping, firefly flashing, and cricket chirping},\ }\href@noop {} {\bibfield
   {journal} {\bibinfo  {journal} {Chaos: An Interdisciplinary Journal of
  Nonlinear Science}\ }\textbf {\bibinfo {volume} {27}},\ \bibinfo {pages}
  {051101} (\bibinfo {year} {2017})}\BibitemShut {NoStop}%
\bibitem [{\citenamefont {Kuramoto}(1975)}]{kuramoto1975international}%
  \BibitemOpen
  \bibfield  {author} {\bibinfo {author} {\bibfnamefont {Y.}~\bibnamefont
  {Kuramoto}},\ }\bibfield  {title} {\bibinfo {title} {International symposium
  on mathematical problems in theoretical physics},\ }\href@noop {} {\bibfield
  {journal} {\bibinfo  {journal} {Lecture notes in Physics}\ }\textbf {\bibinfo
  {volume} {30}},\ \bibinfo {pages} {420} (\bibinfo {year} {1975})}\BibitemShut
  {NoStop}%
\bibitem [{\citenamefont {Sun}\ \emph {et~al.}(2018)\citenamefont {Sun},
  \citenamefont {Yang}, \citenamefont {Gu}, \citenamefont {Chen}, \citenamefont
  {Zhou} \emph {et~al.}}]{sun2018electrochemical}%
  \BibitemOpen
  \bibfield  {author} {\bibinfo {author} {\bibfnamefont {Y.}~\bibnamefont
  {Sun}}, \bibinfo {author} {\bibfnamefont {Z.}~\bibnamefont {Yang}}, \bibinfo
  {author} {\bibfnamefont {L.}~\bibnamefont {Gu}}, \bibinfo {author}
  {\bibfnamefont {Y.}~\bibnamefont {Chen}}, \bibinfo {author} {\bibfnamefont
  {H.}~\bibnamefont {Zhou}}, \emph {et~al.},\ }\bibfield  {title} {\bibinfo
  {title} {Electrochemical oscillation in li-ion batteries},\ }\href@noop {}
  {\bibfield  {journal} {\bibinfo  {journal} {Joule}\ }\textbf {\bibinfo
  {volume} {2}},\ \bibinfo {pages} {1265} (\bibinfo {year} {2018})}\BibitemShut
  {NoStop}%
\bibitem [{\citenamefont {Park}\ \emph {et~al.}(2021)\citenamefont {Park},
  \citenamefont {Zhao}, \citenamefont {Kang}, \citenamefont {Lim},
  \citenamefont {Chen}, \citenamefont {Yu}, \citenamefont {Braatz},
  \citenamefont {Shapiro}, \citenamefont {Hong}, \citenamefont {Toney} \emph
  {et~al.}}]{park2021fictitious}%
  \BibitemOpen
  \bibfield  {author} {\bibinfo {author} {\bibfnamefont {J.}~\bibnamefont
  {Park}}, \bibinfo {author} {\bibfnamefont {H.}~\bibnamefont {Zhao}}, \bibinfo
  {author} {\bibfnamefont {S.~D.}\ \bibnamefont {Kang}}, \bibinfo {author}
  {\bibfnamefont {K.}~\bibnamefont {Lim}}, \bibinfo {author} {\bibfnamefont
  {C.-C.}\ \bibnamefont {Chen}}, \bibinfo {author} {\bibfnamefont {Y.-S.}\
  \bibnamefont {Yu}}, \bibinfo {author} {\bibfnamefont {R.~D.}\ \bibnamefont
  {Braatz}}, \bibinfo {author} {\bibfnamefont {D.~A.}\ \bibnamefont {Shapiro}},
  \bibinfo {author} {\bibfnamefont {J.}~\bibnamefont {Hong}}, \bibinfo {author}
  {\bibfnamefont {M.~F.}\ \bibnamefont {Toney}}, \emph {et~al.},\ }\bibfield
  {title} {\bibinfo {title} {Fictitious phase separation in li layered oxides
  driven by electro-autocatalysis},\ }\href@noop {} {\bibfield  {journal}
  {\bibinfo  {journal} {Nature Materials}\ }\textbf {\bibinfo {volume} {20}},\
  \bibinfo {pages} {991} (\bibinfo {year} {2021})}\BibitemShut {NoStop}%
\bibitem [{\citenamefont {Ferguson}\ and\ \citenamefont
  {Bazant}(2012)}]{ferguson2012nonequilibrium}%
  \BibitemOpen
  \bibfield  {author} {\bibinfo {author} {\bibfnamefont {T.~R.}\ \bibnamefont
  {Ferguson}}\ and\ \bibinfo {author} {\bibfnamefont {M.~Z.}\ \bibnamefont
  {Bazant}},\ }\bibfield  {title} {\bibinfo {title} {Nonequilibrium
  thermodynamics of porous electrodes},\ }\href@noop {} {\bibfield  {journal}
  {\bibinfo  {journal} {Journal of The Electrochemical Society}\ }\textbf
  {\bibinfo {volume} {159}},\ \bibinfo {pages} {A1967} (\bibinfo {year}
  {2012})}\BibitemShut {NoStop}%
\bibitem [{\citenamefont {Ferguson}\ and\ \citenamefont
  {Bazant}(2014)}]{ferguson2014phase}%
  \BibitemOpen
  \bibfield  {author} {\bibinfo {author} {\bibfnamefont {T.~R.}\ \bibnamefont
  {Ferguson}}\ and\ \bibinfo {author} {\bibfnamefont {M.~Z.}\ \bibnamefont
  {Bazant}},\ }\bibfield  {title} {\bibinfo {title} {Phase transformation
  dynamics in porous battery electrodes},\ }\href@noop {} {\bibfield  {journal}
  {\bibinfo  {journal} {Electrochimica Acta}\ }\textbf {\bibinfo {volume}
  {146}},\ \bibinfo {pages} {89} (\bibinfo {year} {2014})}\BibitemShut
  {NoStop}%
\bibitem [{\citenamefont {Li}\ \emph {et~al.}(2014)\citenamefont {Li},
  \citenamefont {El~Gabaly}, \citenamefont {Ferguson}, \citenamefont {Smith},
  \citenamefont {Bartelt}, \citenamefont {Sugar}, \citenamefont {Fenton},
  \citenamefont {Cogswell}, \citenamefont {Kilcoyne}, \citenamefont
  {Tyliszczak} \emph {et~al.}}]{li2014current}%
  \BibitemOpen
  \bibfield  {author} {\bibinfo {author} {\bibfnamefont {Y.}~\bibnamefont
  {Li}}, \bibinfo {author} {\bibfnamefont {F.}~\bibnamefont {El~Gabaly}},
  \bibinfo {author} {\bibfnamefont {T.~R.}\ \bibnamefont {Ferguson}}, \bibinfo
  {author} {\bibfnamefont {R.~B.}\ \bibnamefont {Smith}}, \bibinfo {author}
  {\bibfnamefont {N.~C.}\ \bibnamefont {Bartelt}}, \bibinfo {author}
  {\bibfnamefont {J.~D.}\ \bibnamefont {Sugar}}, \bibinfo {author}
  {\bibfnamefont {K.~R.}\ \bibnamefont {Fenton}}, \bibinfo {author}
  {\bibfnamefont {D.~A.}\ \bibnamefont {Cogswell}}, \bibinfo {author}
  {\bibfnamefont {A.}~\bibnamefont {Kilcoyne}}, \bibinfo {author}
  {\bibfnamefont {T.}~\bibnamefont {Tyliszczak}}, \emph {et~al.},\ }\bibfield
  {title} {\bibinfo {title} {Current-induced transition from
  particle-by-particle to concurrent intercalation in phase-separating battery
  electrodes},\ }\href@noop {} {\bibfield  {journal} {\bibinfo  {journal}
  {Nature materials}\ }\textbf {\bibinfo {volume} {13}},\ \bibinfo {pages}
  {1149} (\bibinfo {year} {2014})}\BibitemShut {NoStop}%
\bibitem [{\citenamefont {Lim}\ \emph {et~al.}(2016)\citenamefont {Lim},
  \citenamefont {Li}, \citenamefont {Alsem}, \citenamefont {So}, \citenamefont
  {Lee}, \citenamefont {Bai}, \citenamefont {Cogswell}, \citenamefont {Liu},
  \citenamefont {Jin}, \citenamefont {Yu} \emph {et~al.}}]{lim2016origin}%
  \BibitemOpen
  \bibfield  {author} {\bibinfo {author} {\bibfnamefont {J.}~\bibnamefont
  {Lim}}, \bibinfo {author} {\bibfnamefont {Y.}~\bibnamefont {Li}}, \bibinfo
  {author} {\bibfnamefont {D.~H.}\ \bibnamefont {Alsem}}, \bibinfo {author}
  {\bibfnamefont {H.}~\bibnamefont {So}}, \bibinfo {author} {\bibfnamefont
  {S.~C.}\ \bibnamefont {Lee}}, \bibinfo {author} {\bibfnamefont
  {P.}~\bibnamefont {Bai}}, \bibinfo {author} {\bibfnamefont {D.~A.}\
  \bibnamefont {Cogswell}}, \bibinfo {author} {\bibfnamefont {X.}~\bibnamefont
  {Liu}}, \bibinfo {author} {\bibfnamefont {N.}~\bibnamefont {Jin}}, \bibinfo
  {author} {\bibfnamefont {Y.-s.}\ \bibnamefont {Yu}}, \emph {et~al.},\
  }\bibfield  {title} {\bibinfo {title} {Origin and hysteresis of lithium
  compositional spatiodynamics within battery primary particles},\ }\href@noop
  {} {\bibfield  {journal} {\bibinfo  {journal} {Science}\ }\textbf {\bibinfo
  {volume} {353}},\ \bibinfo {pages} {566} (\bibinfo {year}
  {2016})}\BibitemShut {NoStop}%
\bibitem [{\citenamefont {Baca{\"e}r}(2011)}]{bacaer2011short}%
  \BibitemOpen
  \bibfield  {author} {\bibinfo {author} {\bibfnamefont {N.}~\bibnamefont
  {Baca{\"e}r}},\ }\href@noop {} {\emph {\bibinfo {title} {A short history of
  mathematical population dynamics}}},\ Vol.\ \bibinfo {volume} {618}\
  (\bibinfo  {publisher} {Springer},\ \bibinfo {year} {2011})\BibitemShut
  {NoStop}%
\bibitem [{\citenamefont {Herrmann}\ \emph {et~al.}(2012)\citenamefont
  {Herrmann}, \citenamefont {Niethammer},\ and\ \citenamefont
  {Vel{\'a}zquez}}]{herrmann2012kramers}%
  \BibitemOpen
  \bibfield  {author} {\bibinfo {author} {\bibfnamefont {M.}~\bibnamefont
  {Herrmann}}, \bibinfo {author} {\bibfnamefont {B.}~\bibnamefont
  {Niethammer}},\ and\ \bibinfo {author} {\bibfnamefont {J.~J.}\ \bibnamefont
  {Vel{\'a}zquez}},\ }\bibfield  {title} {\bibinfo {title} {Kramers and
  non-kramers phase transitions in many-particle systems with dynamical
  constraint},\ }\href@noop {} {\bibfield  {journal} {\bibinfo  {journal}
  {Multiscale Modeling \& Simulation}\ }\textbf {\bibinfo {volume} {10}},\
  \bibinfo {pages} {818} (\bibinfo {year} {2012})}\BibitemShut {NoStop}%
\bibitem [{\citenamefont {Herrmann}\ \emph {et~al.}(2014)\citenamefont
  {Herrmann}, \citenamefont {Niethammer},\ and\ \citenamefont
  {Velazquez}}]{herrmann2014rate}%
  \BibitemOpen
  \bibfield  {author} {\bibinfo {author} {\bibfnamefont {M.}~\bibnamefont
  {Herrmann}}, \bibinfo {author} {\bibfnamefont {B.}~\bibnamefont
  {Niethammer}},\ and\ \bibinfo {author} {\bibfnamefont {J.~J.}\ \bibnamefont
  {Velazquez}},\ }\bibfield  {title} {\bibinfo {title} {Rate-independent
  dynamics and kramers-type phase transitions in nonlocal fokker--planck
  equations with dynamical control},\ }\href@noop {} {\bibfield  {journal}
  {\bibinfo  {journal} {Archive for Rational Mechanics and Analysis}\ }\textbf
  {\bibinfo {volume} {214}},\ \bibinfo {pages} {803} (\bibinfo {year}
  {2014})}\BibitemShut {NoStop}%
\bibitem [{\citenamefont {Kolmogorov}\ \emph {et~al.}()\citenamefont
  {Kolmogorov}, \citenamefont {Petrovskii},\ and\ \citenamefont
  {Piskunov}}]{kolmogorov1study}%
  \BibitemOpen
  \bibfield  {author} {\bibinfo {author} {\bibfnamefont {A.}~\bibnamefont
  {Kolmogorov}}, \bibinfo {author} {\bibfnamefont {I.}~\bibnamefont
  {Petrovskii}},\ and\ \bibinfo {author} {\bibfnamefont {N.}~\bibnamefont
  {Piskunov}},\ }\bibfield  {title} {\bibinfo {title} {A study of the diffusion
  equation with increase in the amount of substance, and its application to a
  biological problem. {\"u}bersetzung aus: Bulletin of the moscow state
  university series a 1: 1-26, 1937},\ }\href@noop {} {\bibfield  {journal}
  {\bibinfo  {journal} {Selected Works of AN Kolmogorov}\ }\textbf {\bibinfo
  {volume} {1}}}\BibitemShut {NoStop}%
\bibitem [{\citenamefont {Fisher}(1958)}]{fisher1958genetical}%
  \BibitemOpen
  \bibfield  {author} {\bibinfo {author} {\bibfnamefont {R.~A.}\ \bibnamefont
  {Fisher}},\ }\href@noop {} {\emph {\bibinfo {title} {The genetical theory of
  natural selection}}}\ (\bibinfo {year} {1958})\BibitemShut {NoStop}%
\bibitem [{\citenamefont {Wright}(1984)}]{wright1984evolution}%
  \BibitemOpen
  \bibfield  {author} {\bibinfo {author} {\bibfnamefont {S.}~\bibnamefont
  {Wright}},\ }\href@noop {} {\emph {\bibinfo {title} {Evolution and the
  Genetics of Populations, Volume 2: Theory of gene frequencies}}},\
  Vol.~\bibinfo {volume} {2}\ (\bibinfo  {publisher} {University of Chicago
  press},\ \bibinfo {year} {1984})\BibitemShut {NoStop}%
\bibitem [{\citenamefont {Desai}\ and\ \citenamefont
  {Fisher}(2007)}]{desai2007beneficial}%
  \BibitemOpen
  \bibfield  {author} {\bibinfo {author} {\bibfnamefont {M.~M.}\ \bibnamefont
  {Desai}}\ and\ \bibinfo {author} {\bibfnamefont {D.~S.}\ \bibnamefont
  {Fisher}},\ }\bibfield  {title} {\bibinfo {title} {Beneficial
  mutation--selection balance and the effect of linkage on positive
  selection},\ }\href@noop {} {\bibfield  {journal} {\bibinfo  {journal}
  {Genetics}\ }\textbf {\bibinfo {volume} {176}},\ \bibinfo {pages} {1759}
  (\bibinfo {year} {2007})}\BibitemShut {NoStop}%
\bibitem [{\citenamefont {Mustonen}\ and\ \citenamefont
  {L{\"a}ssig}(2009)}]{mustonen2009fitness}%
  \BibitemOpen
  \bibfield  {author} {\bibinfo {author} {\bibfnamefont {V.}~\bibnamefont
  {Mustonen}}\ and\ \bibinfo {author} {\bibfnamefont {M.}~\bibnamefont
  {L{\"a}ssig}},\ }\bibfield  {title} {\bibinfo {title} {From fitness
  landscapes to seascapes: non-equilibrium dynamics of selection and
  adaptation},\ }\href@noop {} {\bibfield  {journal} {\bibinfo  {journal}
  {Trends in genetics}\ }\textbf {\bibinfo {volume} {25}},\ \bibinfo {pages}
  {111} (\bibinfo {year} {2009})}\BibitemShut {NoStop}%
\bibitem [{\citenamefont {Guha}\ \emph {et~al.}(2017)\citenamefont {Guha},
  \citenamefont {Patra},\ and\ \citenamefont {Vaisakh}}]{guha2017remaining}%
  \BibitemOpen
  \bibfield  {author} {\bibinfo {author} {\bibfnamefont {A.}~\bibnamefont
  {Guha}}, \bibinfo {author} {\bibfnamefont {A.}~\bibnamefont {Patra}},\ and\
  \bibinfo {author} {\bibfnamefont {K.}~\bibnamefont {Vaisakh}},\ }\bibfield
  {title} {\bibinfo {title} {Remaining useful life estimation of lithium-ion
  batteries based on the internal resistance growth model},\ }in\ \href@noop {}
  {\emph {\bibinfo {booktitle} {2017 Indian Control Conference (ICC)}}}\
  (\bibinfo {organization} {IEEE},\ \bibinfo {year} {2017})\ pp.\ \bibinfo
  {pages} {33--38}\BibitemShut {NoStop}%
\bibitem [{\citenamefont {Kauffman}\ and\ \citenamefont
  {Johnsen}(1991)}]{kauffman1991coevolution}%
  \BibitemOpen
  \bibfield  {author} {\bibinfo {author} {\bibfnamefont {S.~A.}\ \bibnamefont
  {Kauffman}}\ and\ \bibinfo {author} {\bibfnamefont {S.}~\bibnamefont
  {Johnsen}},\ }\bibfield  {title} {\bibinfo {title} {Coevolution to the edge
  of chaos: coupled fitness landscapes, poised states, and coevolutionary
  avalanches},\ }\href@noop {} {\bibfield  {journal} {\bibinfo  {journal}
  {Journal of theoretical biology}\ }\textbf {\bibinfo {volume} {149}},\
  \bibinfo {pages} {467} (\bibinfo {year} {1991})}\BibitemShut {NoStop}%
\bibitem [{\citenamefont {Newman}\ and\ \citenamefont
  {Thomas-Alyea}(2012)}]{newman2012electrochemical}%
  \BibitemOpen
  \bibfield  {author} {\bibinfo {author} {\bibfnamefont {J.}~\bibnamefont
  {Newman}}\ and\ \bibinfo {author} {\bibfnamefont {K.~E.}\ \bibnamefont
  {Thomas-Alyea}},\ }\href@noop {} {\emph {\bibinfo {title} {Electrochemical
  systems}}}\ (\bibinfo  {publisher} {John Wiley \& Sons},\ \bibinfo {year}
  {2012})\BibitemShut {NoStop}%
\bibitem [{\citenamefont {Doyle}\ \emph {et~al.}(1993)\citenamefont {Doyle},
  \citenamefont {Fuller},\ and\ \citenamefont {Newman}}]{doyle1993modeling}%
  \BibitemOpen
  \bibfield  {author} {\bibinfo {author} {\bibfnamefont {M.}~\bibnamefont
  {Doyle}}, \bibinfo {author} {\bibfnamefont {T.~F.}\ \bibnamefont {Fuller}},\
  and\ \bibinfo {author} {\bibfnamefont {J.}~\bibnamefont {Newman}},\
  }\bibfield  {title} {\bibinfo {title} {Modeling of galvanostatic charge and
  discharge of the lithium/polymer/insertion cell},\ }\href@noop {} {\bibfield
  {journal} {\bibinfo  {journal} {Journal of the Electrochemical society}\
  }\textbf {\bibinfo {volume} {140}},\ \bibinfo {pages} {1526} (\bibinfo {year}
  {1993})}\BibitemShut {NoStop}%
\bibitem [{\citenamefont {Ramkrishna}(2000)}]{ramkrishna2000population}%
  \BibitemOpen
  \bibfield  {author} {\bibinfo {author} {\bibfnamefont {D.}~\bibnamefont
  {Ramkrishna}},\ }\href@noop {} {\emph {\bibinfo {title} {Population balances:
  Theory and applications to particulate systems in engineering}}}\ (\bibinfo
  {publisher} {Elsevier},\ \bibinfo {year} {2000})\BibitemShut {NoStop}%
\bibitem [{\citenamefont {Ramkrishna}\ and\ \citenamefont
  {Singh}(2014)}]{ramkrishna2014population}%
  \BibitemOpen
  \bibfield  {author} {\bibinfo {author} {\bibfnamefont {D.}~\bibnamefont
  {Ramkrishna}}\ and\ \bibinfo {author} {\bibfnamefont {M.~R.}\ \bibnamefont
  {Singh}},\ }\bibfield  {title} {\bibinfo {title} {Population balance
  modeling: current status and future prospects},\ }\href@noop {} {\bibfield
  {journal} {\bibinfo  {journal} {Annual review of chemical and biomolecular
  engineering}\ }\textbf {\bibinfo {volume} {5}},\ \bibinfo {pages} {123}
  (\bibinfo {year} {2014})}\BibitemShut {NoStop}%
\bibitem [{\citenamefont {Frank}(2005)}]{frank2005nonlinear}%
  \BibitemOpen
  \bibfield  {author} {\bibinfo {author} {\bibfnamefont {T.~D.}\ \bibnamefont
  {Frank}},\ }\href@noop {} {\emph {\bibinfo {title} {Nonlinear Fokker-Planck
  equations: fundamentals and applications}}}\ (\bibinfo  {publisher} {Springer
  Science \& Business Media},\ \bibinfo {year} {2005})\BibitemShut {NoStop}%
\bibitem [{\citenamefont {Zhao}\ and\ \citenamefont
  {Bazant}(2019)}]{zhao2019population}%
  \BibitemOpen
  \bibfield  {author} {\bibinfo {author} {\bibfnamefont {H.}~\bibnamefont
  {Zhao}}\ and\ \bibinfo {author} {\bibfnamefont {M.~Z.}\ \bibnamefont
  {Bazant}},\ }\bibfield  {title} {\bibinfo {title} {Population dynamics of
  driven autocatalytic reactive mixtures},\ }\href@noop {} {\bibfield
  {journal} {\bibinfo  {journal} {Physical Review E}\ }\textbf {\bibinfo
  {volume} {100}},\ \bibinfo {pages} {012144} (\bibinfo {year}
  {2019})}\BibitemShut {NoStop}%
\bibitem [{\citenamefont {Dreyer}\ \emph {et~al.}(2011)\citenamefont {Dreyer},
  \citenamefont {Guhlke},\ and\ \citenamefont {Huth}}]{dreyer2011behavior}%
  \BibitemOpen
  \bibfield  {author} {\bibinfo {author} {\bibfnamefont {W.}~\bibnamefont
  {Dreyer}}, \bibinfo {author} {\bibfnamefont {C.}~\bibnamefont {Guhlke}},\
  and\ \bibinfo {author} {\bibfnamefont {R.}~\bibnamefont {Huth}},\ }\bibfield
  {title} {\bibinfo {title} {The behavior of a many-particle electrode in a
  lithium-ion battery},\ }\href@noop {} {\bibfield  {journal} {\bibinfo
  {journal} {Physica D: Nonlinear Phenomena}\ }\textbf {\bibinfo {volume}
  {240}},\ \bibinfo {pages} {1008} (\bibinfo {year} {2011})}\BibitemShut
  {NoStop}%
\bibitem [{\citenamefont {Yang}\ \emph {et~al.}(2019)\citenamefont {Yang},
  \citenamefont {Xu}, \citenamefont {Zhang}, \citenamefont {Lee}, \citenamefont
  {Mu}, \citenamefont {Liu}, \citenamefont {Waters}, \citenamefont {Spence},
  \citenamefont {Xu}, \citenamefont {Wei} \emph
  {et~al.}}]{yang2019quantification}%
  \BibitemOpen
  \bibfield  {author} {\bibinfo {author} {\bibfnamefont {Y.}~\bibnamefont
  {Yang}}, \bibinfo {author} {\bibfnamefont {R.}~\bibnamefont {Xu}}, \bibinfo
  {author} {\bibfnamefont {K.}~\bibnamefont {Zhang}}, \bibinfo {author}
  {\bibfnamefont {S.-J.}\ \bibnamefont {Lee}}, \bibinfo {author} {\bibfnamefont
  {L.}~\bibnamefont {Mu}}, \bibinfo {author} {\bibfnamefont {P.}~\bibnamefont
  {Liu}}, \bibinfo {author} {\bibfnamefont {C.~K.}\ \bibnamefont {Waters}},
  \bibinfo {author} {\bibfnamefont {S.}~\bibnamefont {Spence}}, \bibinfo
  {author} {\bibfnamefont {Z.}~\bibnamefont {Xu}}, \bibinfo {author}
  {\bibfnamefont {C.}~\bibnamefont {Wei}}, \emph {et~al.},\ }\bibfield  {title}
  {\bibinfo {title} {Quantification of heterogeneous degradation in li-ion
  batteries},\ }\href@noop {} {\bibfield  {journal} {\bibinfo  {journal}
  {Advanced Energy Materials}\ }\textbf {\bibinfo {volume} {9}},\ \bibinfo
  {pages} {1900674} (\bibinfo {year} {2019})}\BibitemShut {NoStop}%
\bibitem [{\citenamefont {Xu}\ \emph {et~al.}(2019)\citenamefont {Xu},
  \citenamefont {Yang}, \citenamefont {Yin}, \citenamefont {Liu}, \citenamefont
  {Cloetens}, \citenamefont {Liu}, \citenamefont {Lin},\ and\ \citenamefont
  {Zhao}}]{xu2019heterogeneous}%
  \BibitemOpen
  \bibfield  {author} {\bibinfo {author} {\bibfnamefont {R.}~\bibnamefont
  {Xu}}, \bibinfo {author} {\bibfnamefont {Y.}~\bibnamefont {Yang}}, \bibinfo
  {author} {\bibfnamefont {F.}~\bibnamefont {Yin}}, \bibinfo {author}
  {\bibfnamefont {P.}~\bibnamefont {Liu}}, \bibinfo {author} {\bibfnamefont
  {P.}~\bibnamefont {Cloetens}}, \bibinfo {author} {\bibfnamefont
  {Y.}~\bibnamefont {Liu}}, \bibinfo {author} {\bibfnamefont {F.}~\bibnamefont
  {Lin}},\ and\ \bibinfo {author} {\bibfnamefont {K.}~\bibnamefont {Zhao}},\
  }\bibfield  {title} {\bibinfo {title} {Heterogeneous damage in li-ion
  batteries: Experimental analysis and theoretical modeling},\ }\href@noop {}
  {\bibfield  {journal} {\bibinfo  {journal} {Journal of the Mechanics and
  Physics of Solids}\ }\textbf {\bibinfo {volume} {129}},\ \bibinfo {pages}
  {160} (\bibinfo {year} {2019})}\BibitemShut {NoStop}%
\bibitem [{\citenamefont {Faulkner}\ and\ \citenamefont
  {Bard}(2002)}]{faulkner2002electrochemical}%
  \BibitemOpen
  \bibfield  {author} {\bibinfo {author} {\bibfnamefont {L.~R.}\ \bibnamefont
  {Faulkner}}\ and\ \bibinfo {author} {\bibfnamefont {A.~J.}\ \bibnamefont
  {Bard}},\ }\href@noop {} {\emph {\bibinfo {title} {Electrochemical methods:
  fundamentals and applications}}}\ (\bibinfo  {publisher} {John Wiley and
  Sons},\ \bibinfo {year} {2002})\BibitemShut {NoStop}%
\bibitem [{\citenamefont {Fraggedakis}\ \emph {et~al.}(2021)\citenamefont
  {Fraggedakis}, \citenamefont {McEldrew}, \citenamefont {Smith}, \citenamefont
  {Krishnan}, \citenamefont {Zhang}, \citenamefont {Bai}, \citenamefont
  {Chueh}, \citenamefont {Shao-Horn},\ and\ \citenamefont
  {Bazant}}]{fraggedakis2021theory}%
  \BibitemOpen
  \bibfield  {author} {\bibinfo {author} {\bibfnamefont {D.}~\bibnamefont
  {Fraggedakis}}, \bibinfo {author} {\bibfnamefont {M.}~\bibnamefont
  {McEldrew}}, \bibinfo {author} {\bibfnamefont {R.~B.}\ \bibnamefont {Smith}},
  \bibinfo {author} {\bibfnamefont {Y.}~\bibnamefont {Krishnan}}, \bibinfo
  {author} {\bibfnamefont {Y.}~\bibnamefont {Zhang}}, \bibinfo {author}
  {\bibfnamefont {P.}~\bibnamefont {Bai}}, \bibinfo {author} {\bibfnamefont
  {W.~C.}\ \bibnamefont {Chueh}}, \bibinfo {author} {\bibfnamefont
  {Y.}~\bibnamefont {Shao-Horn}},\ and\ \bibinfo {author} {\bibfnamefont
  {M.~Z.}\ \bibnamefont {Bazant}},\ }\bibfield  {title} {\bibinfo {title}
  {Theory of coupled ion-electron transfer kinetics},\ }\href@noop {}
  {\bibfield  {journal} {\bibinfo  {journal} {Electrochimica Acta}\ }\textbf
  {\bibinfo {volume} {367}},\ \bibinfo {pages} {137432} (\bibinfo {year}
  {2021})}\BibitemShut {NoStop}%
\bibitem [{\citenamefont {Zhang}\ \emph {et~al.}()\citenamefont {Zhang},
  \citenamefont {Fraggedakis}, \citenamefont {Gao}, \citenamefont {Zhuang},
  \citenamefont {Zhu}, \citenamefont {Huang}, \citenamefont {Eisenach},
  \citenamefont {Giordano}, \citenamefont {Tatara}, \citenamefont {Stephens},
  \citenamefont {Bazant},\ and\ \citenamefont
  {Shao-Horn}}]{Zhang2022_CIET_preprint}%
  \BibitemOpen
  \bibfield  {author} {\bibinfo {author} {\bibfnamefont {Y.}~\bibnamefont
  {Zhang}}, \bibinfo {author} {\bibfnamefont {D.}~\bibnamefont {Fraggedakis}},
  \bibinfo {author} {\bibfnamefont {T.}~\bibnamefont {Gao}}, \bibinfo {author}
  {\bibfnamefont {D.}~\bibnamefont {Zhuang}}, \bibinfo {author} {\bibfnamefont
  {Y.~G.}\ \bibnamefont {Zhu}}, \bibinfo {author} {\bibfnamefont
  {B.}~\bibnamefont {Huang}}, \bibinfo {author} {\bibfnamefont
  {R.}~\bibnamefont {Eisenach}}, \bibinfo {author} {\bibfnamefont
  {L.}~\bibnamefont {Giordano}}, \bibinfo {author} {\bibfnamefont
  {R.}~\bibnamefont {Tatara}}, \bibinfo {author} {\bibfnamefont {R.~M.}\
  \bibnamefont {Stephens}}, \bibinfo {author} {\bibfnamefont {M.~Z.}\
  \bibnamefont {Bazant}},\ and\ \bibinfo {author} {\bibfnamefont
  {Y.}~\bibnamefont {Shao-Horn}},\ }\bibfield  {title} {\bibinfo {title}
  {Lithium intercalation by coupled ion-electron transfer},\ }\bibinfo {note}
  {in preparation}\BibitemShut {NoStop}%
\bibitem [{\citenamefont {Zeng}\ \emph {et~al.}(2014)\citenamefont {Zeng},
  \citenamefont {Smith}, \citenamefont {Bai},\ and\ \citenamefont
  {Bazant}}]{zeng2014simple}%
  \BibitemOpen
  \bibfield  {author} {\bibinfo {author} {\bibfnamefont {Y.}~\bibnamefont
  {Zeng}}, \bibinfo {author} {\bibfnamefont {R.~B.}\ \bibnamefont {Smith}},
  \bibinfo {author} {\bibfnamefont {P.}~\bibnamefont {Bai}},\ and\ \bibinfo
  {author} {\bibfnamefont {M.~Z.}\ \bibnamefont {Bazant}},\ }\bibfield  {title}
  {\bibinfo {title} {Simple formula for marcus--hush--chidsey kinetics},\
  }\href@noop {} {\bibfield  {journal} {\bibinfo  {journal} {Journal of
  Electroanalytical Chemistry}\ }\textbf {\bibinfo {volume} {735}},\ \bibinfo
  {pages} {77} (\bibinfo {year} {2014})}\BibitemShut {NoStop}%
\bibitem [{\citenamefont {Bazant}(2017)}]{bazant2017thermodynamic}%
  \BibitemOpen
  \bibfield  {author} {\bibinfo {author} {\bibfnamefont {M.~Z.}\ \bibnamefont
  {Bazant}},\ }\bibfield  {title} {\bibinfo {title} {Thermodynamic stability of
  driven open systems and control of phase separation by
  electro-autocatalysis},\ }\href@noop {} {\bibfield  {journal} {\bibinfo
  {journal} {Faraday discussions}\ }\textbf {\bibinfo {volume} {199}},\
  \bibinfo {pages} {423} (\bibinfo {year} {2017})}\BibitemShut {NoStop}%
\bibitem [{\citenamefont {Bazant}(2013)}]{bazant2013theory}%
  \BibitemOpen
  \bibfield  {author} {\bibinfo {author} {\bibfnamefont {M.~Z.}\ \bibnamefont
  {Bazant}},\ }\bibfield  {title} {\bibinfo {title} {Theory of chemical
  kinetics and charge transfer based on nonequilibrium thermodynamics},\
  }\href@noop {} {\bibfield  {journal} {\bibinfo  {journal} {Accounts of
  chemical research}\ }\textbf {\bibinfo {volume} {46}},\ \bibinfo {pages}
  {1144} (\bibinfo {year} {2013})}\BibitemShut {NoStop}%
\bibitem [{\citenamefont {Miller}\ \emph {et~al.}(1984)\citenamefont {Miller},
  \citenamefont {Calcaterra},\ and\ \citenamefont
  {Closs}}]{miller1984intramolecular}%
  \BibitemOpen
  \bibfield  {author} {\bibinfo {author} {\bibfnamefont {J.~R.}\ \bibnamefont
  {Miller}}, \bibinfo {author} {\bibfnamefont {L.}~\bibnamefont {Calcaterra}},\
  and\ \bibinfo {author} {\bibfnamefont {G.}~\bibnamefont {Closs}},\ }\bibfield
   {title} {\bibinfo {title} {Intramolecular long-distance electron transfer in
  radical anions. the effects of free energy and solvent on the reaction
  rates},\ }\href@noop {} {\bibfield  {journal} {\bibinfo  {journal} {Journal
  of the American Chemical Society}\ }\textbf {\bibinfo {volume} {106}},\
  \bibinfo {pages} {3047} (\bibinfo {year} {1984})}\BibitemShut {NoStop}%
\bibitem [{\citenamefont {Chidsey}(1991)}]{chidsey1991free}%
  \BibitemOpen
  \bibfield  {author} {\bibinfo {author} {\bibfnamefont {C.~E.}\ \bibnamefont
  {Chidsey}},\ }\bibfield  {title} {\bibinfo {title} {Free energy and
  temperature dependence of electron transfer at the metal-electrolyte
  interface},\ }\href@noop {} {\bibfield  {journal} {\bibinfo  {journal}
  {Science}\ }\textbf {\bibinfo {volume} {251}},\ \bibinfo {pages} {919}
  (\bibinfo {year} {1991})}\BibitemShut {NoStop}%
\bibitem [{\citenamefont {Pinson}\ and\ \citenamefont
  {Bazant}(2012)}]{pinson2012theory}%
  \BibitemOpen
  \bibfield  {author} {\bibinfo {author} {\bibfnamefont {M.~B.}\ \bibnamefont
  {Pinson}}\ and\ \bibinfo {author} {\bibfnamefont {M.~Z.}\ \bibnamefont
  {Bazant}},\ }\bibfield  {title} {\bibinfo {title} {Theory of sei formation in
  rechargeable batteries: capacity fade, accelerated aging and lifetime
  prediction},\ }\href@noop {} {\bibfield  {journal} {\bibinfo  {journal}
  {Journal of the Electrochemical Society}\ }\textbf {\bibinfo {volume}
  {160}},\ \bibinfo {pages} {A243} (\bibinfo {year} {2012})}\BibitemShut
  {NoStop}%
\bibitem [{\citenamefont {Nie}\ \emph {et~al.}(2013)\citenamefont {Nie},
  \citenamefont {Abraham}, \citenamefont {Seo}, \citenamefont {Chen},
  \citenamefont {Bose},\ and\ \citenamefont {Lucht}}]{nie2013role}%
  \BibitemOpen
  \bibfield  {author} {\bibinfo {author} {\bibfnamefont {M.}~\bibnamefont
  {Nie}}, \bibinfo {author} {\bibfnamefont {D.~P.}\ \bibnamefont {Abraham}},
  \bibinfo {author} {\bibfnamefont {D.~M.}\ \bibnamefont {Seo}}, \bibinfo
  {author} {\bibfnamefont {Y.}~\bibnamefont {Chen}}, \bibinfo {author}
  {\bibfnamefont {A.}~\bibnamefont {Bose}},\ and\ \bibinfo {author}
  {\bibfnamefont {B.~L.}\ \bibnamefont {Lucht}},\ }\bibfield  {title} {\bibinfo
  {title} {Role of solution structure in solid electrolyte interphase formation
  on graphite with lipf6 in propylene carbonate},\ }\href@noop {} {\bibfield
  {journal} {\bibinfo  {journal} {The Journal of Physical Chemistry C}\
  }\textbf {\bibinfo {volume} {117}},\ \bibinfo {pages} {25381} (\bibinfo
  {year} {2013})}\BibitemShut {NoStop}%
\bibitem [{\citenamefont {Darling}\ and\ \citenamefont
  {Newman}(1997)}]{darling1997modeling}%
  \BibitemOpen
  \bibfield  {author} {\bibinfo {author} {\bibfnamefont {R.}~\bibnamefont
  {Darling}}\ and\ \bibinfo {author} {\bibfnamefont {J.}~\bibnamefont
  {Newman}},\ }\bibfield  {title} {\bibinfo {title} {Modeling a porous
  intercalation electrode with two characteristic particle sizes},\ }\href@noop
  {} {\bibfield  {journal} {\bibinfo  {journal} {Journal of The Electrochemical
  Society}\ }\textbf {\bibinfo {volume} {144}},\ \bibinfo {pages} {4201}
  (\bibinfo {year} {1997})}\BibitemShut {NoStop}%
\bibitem [{\citenamefont {Fuller}\ \emph {et~al.}(1994)\citenamefont {Fuller},
  \citenamefont {Doyle},\ and\ \citenamefont {Newman}}]{fuller1994simulation}%
  \BibitemOpen
  \bibfield  {author} {\bibinfo {author} {\bibfnamefont {T.~F.}\ \bibnamefont
  {Fuller}}, \bibinfo {author} {\bibfnamefont {M.}~\bibnamefont {Doyle}},\ and\
  \bibinfo {author} {\bibfnamefont {J.}~\bibnamefont {Newman}},\ }\bibfield
  {title} {\bibinfo {title} {Simulation and optimization of the dual lithium
  ion insertion cell},\ }\href@noop {} {\bibfield  {journal} {\bibinfo
  {journal} {Journal of the Electrochemical Society}\ }\textbf {\bibinfo
  {volume} {141}},\ \bibinfo {pages} {1} (\bibinfo {year} {1994})}\BibitemShut
  {NoStop}%
\bibitem [{\citenamefont {Smith}\ and\ \citenamefont
  {Bazant}(2017)}]{smith2017multiphase}%
  \BibitemOpen
  \bibfield  {author} {\bibinfo {author} {\bibfnamefont {R.~B.}\ \bibnamefont
  {Smith}}\ and\ \bibinfo {author} {\bibfnamefont {M.~Z.}\ \bibnamefont
  {Bazant}},\ }\bibfield  {title} {\bibinfo {title} {Multiphase porous
  electrode theory},\ }\href@noop {} {\bibfield  {journal} {\bibinfo  {journal}
  {Journal of The Electrochemical Society}\ }\textbf {\bibinfo {volume}
  {164}},\ \bibinfo {pages} {E3291} (\bibinfo {year} {2017})}\BibitemShut
  {NoStop}%
\bibitem [{\citenamefont {Fraggedakis}\ \emph {et~al.}(2020)\citenamefont
  {Fraggedakis}, \citenamefont {Nadkarni}, \citenamefont {Gao}, \citenamefont
  {Zhou}, \citenamefont {Zhang}, \citenamefont {Han}, \citenamefont {Stephens},
  \citenamefont {Shao-Horn},\ and\ \citenamefont
  {Bazant}}]{fraggedakis2020scaling}%
  \BibitemOpen
  \bibfield  {author} {\bibinfo {author} {\bibfnamefont {D.}~\bibnamefont
  {Fraggedakis}}, \bibinfo {author} {\bibfnamefont {N.}~\bibnamefont
  {Nadkarni}}, \bibinfo {author} {\bibfnamefont {T.}~\bibnamefont {Gao}},
  \bibinfo {author} {\bibfnamefont {T.}~\bibnamefont {Zhou}}, \bibinfo {author}
  {\bibfnamefont {Y.}~\bibnamefont {Zhang}}, \bibinfo {author} {\bibfnamefont
  {Y.}~\bibnamefont {Han}}, \bibinfo {author} {\bibfnamefont {R.~M.}\
  \bibnamefont {Stephens}}, \bibinfo {author} {\bibfnamefont {Y.}~\bibnamefont
  {Shao-Horn}},\ and\ \bibinfo {author} {\bibfnamefont {M.~Z.}\ \bibnamefont
  {Bazant}},\ }\bibfield  {title} {\bibinfo {title} {A scaling law to determine
  phase morphologies during ion intercalation},\ }\href@noop {} {\bibfield
  {journal} {\bibinfo  {journal} {Energy \& Environmental Science}\ }\textbf
  {\bibinfo {volume} {13}},\ \bibinfo {pages} {2142} (\bibinfo {year}
  {2020})}\BibitemShut {NoStop}%
\bibitem [{\citenamefont {Li}\ \emph {et~al.}(2019)\citenamefont {Li},
  \citenamefont {Wang}, \citenamefont {Wang}, \citenamefont {Sun},
  \citenamefont {Zhang}, \citenamefont {Yu},\ and\ \citenamefont
  {Li}}]{li2019investigations}%
  \BibitemOpen
  \bibfield  {author} {\bibinfo {author} {\bibfnamefont {Q.}~\bibnamefont
  {Li}}, \bibinfo {author} {\bibfnamefont {Y.}~\bibnamefont {Wang}}, \bibinfo
  {author} {\bibfnamefont {X.}~\bibnamefont {Wang}}, \bibinfo {author}
  {\bibfnamefont {X.}~\bibnamefont {Sun}}, \bibinfo {author} {\bibfnamefont
  {J.-N.}\ \bibnamefont {Zhang}}, \bibinfo {author} {\bibfnamefont
  {X.}~\bibnamefont {Yu}},\ and\ \bibinfo {author} {\bibfnamefont
  {H.}~\bibnamefont {Li}},\ }\bibfield  {title} {\bibinfo {title}
  {Investigations on the fundamental process of cathode electrolyte interphase
  formation and evolution of high-voltage cathodes},\ }\href@noop {} {\bibfield
   {journal} {\bibinfo  {journal} {ACS applied materials \& interfaces}\
  }\textbf {\bibinfo {volume} {12}},\ \bibinfo {pages} {2319} (\bibinfo {year}
  {2019})}\BibitemShut {NoStop}%
\bibitem [{\citenamefont {Mohanty}\ \emph {et~al.}(2016)\citenamefont
  {Mohanty}, \citenamefont {Dahlberg}, \citenamefont {King}, \citenamefont
  {David}, \citenamefont {Sefat}, \citenamefont {Wood}, \citenamefont {Daniel},
  \citenamefont {Dhar}, \citenamefont {Mahajan}, \citenamefont {Lee} \emph
  {et~al.}}]{mohanty2016modification}%
  \BibitemOpen
  \bibfield  {author} {\bibinfo {author} {\bibfnamefont {D.}~\bibnamefont
  {Mohanty}}, \bibinfo {author} {\bibfnamefont {K.}~\bibnamefont {Dahlberg}},
  \bibinfo {author} {\bibfnamefont {D.~M.}\ \bibnamefont {King}}, \bibinfo
  {author} {\bibfnamefont {L.~A.}\ \bibnamefont {David}}, \bibinfo {author}
  {\bibfnamefont {A.~S.}\ \bibnamefont {Sefat}}, \bibinfo {author}
  {\bibfnamefont {D.~L.}\ \bibnamefont {Wood}}, \bibinfo {author}
  {\bibfnamefont {C.}~\bibnamefont {Daniel}}, \bibinfo {author} {\bibfnamefont
  {S.}~\bibnamefont {Dhar}}, \bibinfo {author} {\bibfnamefont {V.}~\bibnamefont
  {Mahajan}}, \bibinfo {author} {\bibfnamefont {M.}~\bibnamefont {Lee}}, \emph
  {et~al.},\ }\bibfield  {title} {\bibinfo {title} {Modification of ni-rich fcg
  nmc and nca cathodes by atomic layer deposition: preventing surface phase
  transitions for high-voltage lithium-ion batteries},\ }\href@noop {}
  {\bibfield  {journal} {\bibinfo  {journal} {Scientific reports}\ }\textbf
  {\bibinfo {volume} {6}},\ \bibinfo {pages} {1} (\bibinfo {year}
  {2016})}\BibitemShut {NoStop}%
\bibitem [{\citenamefont {Li}\ \emph {et~al.}(2020{\natexlab{a}})\citenamefont
  {Li}, \citenamefont {Yao}, \citenamefont {Zheng}, \citenamefont {Fu},
  \citenamefont {Cen}, \citenamefont {Hwang}, \citenamefont {Jin},
  \citenamefont {Orlov}, \citenamefont {Gu}, \citenamefont {Wang} \emph
  {et~al.}}]{li2020direct}%
  \BibitemOpen
  \bibfield  {author} {\bibinfo {author} {\bibfnamefont {S.}~\bibnamefont
  {Li}}, \bibinfo {author} {\bibfnamefont {Z.}~\bibnamefont {Yao}}, \bibinfo
  {author} {\bibfnamefont {J.}~\bibnamefont {Zheng}}, \bibinfo {author}
  {\bibfnamefont {M.}~\bibnamefont {Fu}}, \bibinfo {author} {\bibfnamefont
  {J.}~\bibnamefont {Cen}}, \bibinfo {author} {\bibfnamefont {S.}~\bibnamefont
  {Hwang}}, \bibinfo {author} {\bibfnamefont {H.}~\bibnamefont {Jin}}, \bibinfo
  {author} {\bibfnamefont {A.}~\bibnamefont {Orlov}}, \bibinfo {author}
  {\bibfnamefont {L.}~\bibnamefont {Gu}}, \bibinfo {author} {\bibfnamefont
  {S.}~\bibnamefont {Wang}}, \emph {et~al.},\ }\bibfield  {title} {\bibinfo
  {title} {Direct observation of defect-aided structural evolution in a
  nickel-rich layered cathode},\ }\href@noop {} {\bibfield  {journal} {\bibinfo
   {journal} {Angewandte Chemie International Edition}\ }\textbf {\bibinfo
  {volume} {59}},\ \bibinfo {pages} {22092} (\bibinfo {year}
  {2020}{\natexlab{a}})}\BibitemShut {NoStop}%
\bibitem [{\citenamefont {Li}\ \emph {et~al.}(2020{\natexlab{b}})\citenamefont
  {Li}, \citenamefont {Yuan}, \citenamefont {Zhang}, \citenamefont {Song},
  \citenamefont {Shi},\ and\ \citenamefont {Bock}}]{li2020degradation}%
  \BibitemOpen
  \bibfield  {author} {\bibinfo {author} {\bibfnamefont {T.}~\bibnamefont
  {Li}}, \bibinfo {author} {\bibfnamefont {X.-Z.}\ \bibnamefont {Yuan}},
  \bibinfo {author} {\bibfnamefont {L.}~\bibnamefont {Zhang}}, \bibinfo
  {author} {\bibfnamefont {D.}~\bibnamefont {Song}}, \bibinfo {author}
  {\bibfnamefont {K.}~\bibnamefont {Shi}},\ and\ \bibinfo {author}
  {\bibfnamefont {C.}~\bibnamefont {Bock}},\ }\bibfield  {title} {\bibinfo
  {title} {Degradation mechanisms and mitigation strategies of nickel-rich
  nmc-based lithium-ion batteries},\ }\href@noop {} {\bibfield  {journal}
  {\bibinfo  {journal} {Electrochemical Energy Reviews}\ }\textbf {\bibinfo
  {volume} {3}},\ \bibinfo {pages} {43} (\bibinfo {year}
  {2020}{\natexlab{b}})}\BibitemShut {NoStop}%
\bibitem [{\citenamefont {Manthiram}\ \emph {et~al.}(2016)\citenamefont
  {Manthiram}, \citenamefont {Knight}, \citenamefont {Myung}, \citenamefont
  {Oh},\ and\ \citenamefont {Sun}}]{manthiram2016nickel}%
  \BibitemOpen
  \bibfield  {author} {\bibinfo {author} {\bibfnamefont {A.}~\bibnamefont
  {Manthiram}}, \bibinfo {author} {\bibfnamefont {J.~C.}\ \bibnamefont
  {Knight}}, \bibinfo {author} {\bibfnamefont {S.-T.}\ \bibnamefont {Myung}},
  \bibinfo {author} {\bibfnamefont {S.-M.}\ \bibnamefont {Oh}},\ and\ \bibinfo
  {author} {\bibfnamefont {Y.-K.}\ \bibnamefont {Sun}},\ }\bibfield  {title}
  {\bibinfo {title} {Nickel-rich and lithium-rich layered oxide cathodes:
  progress and perspectives},\ }\href@noop {} {\bibfield  {journal} {\bibinfo
  {journal} {Advanced Energy Materials}\ }\textbf {\bibinfo {volume} {6}},\
  \bibinfo {pages} {1501010} (\bibinfo {year} {2016})}\BibitemShut {NoStop}%
\bibitem [{\citenamefont {Edstroem}\ \emph {et~al.}(2004)\citenamefont
  {Edstroem}, \citenamefont {Gustafsson},\ and\ \citenamefont
  {Thomas}}]{edstroem2004cathode}%
  \BibitemOpen
  \bibfield  {author} {\bibinfo {author} {\bibfnamefont {K.}~\bibnamefont
  {Edstroem}}, \bibinfo {author} {\bibfnamefont {T.}~\bibnamefont
  {Gustafsson}},\ and\ \bibinfo {author} {\bibfnamefont {J.~O.}\ \bibnamefont
  {Thomas}},\ }\bibfield  {title} {\bibinfo {title} {The cathode--electrolyte
  interface in the li-ion battery},\ }\href@noop {} {\bibfield  {journal}
  {\bibinfo  {journal} {Electrochimica Acta}\ }\textbf {\bibinfo {volume}
  {50}},\ \bibinfo {pages} {397} (\bibinfo {year} {2004})}\BibitemShut
  {NoStop}%
\bibitem [{\citenamefont {Maleki Kheimeh~Sari}\ and\ \citenamefont
  {Li}(2019)}]{maleki2019controllable}%
  \BibitemOpen
  \bibfield  {author} {\bibinfo {author} {\bibfnamefont {H.}~\bibnamefont
  {Maleki Kheimeh~Sari}}\ and\ \bibinfo {author} {\bibfnamefont
  {X.}~\bibnamefont {Li}},\ }\bibfield  {title} {\bibinfo {title} {Controllable
  cathode--electrolyte interface of li [ni0. 8co0. 1mn0. 1] o2 for lithium ion
  batteries: a review},\ }\href@noop {} {\bibfield  {journal} {\bibinfo
  {journal} {Advanced Energy Materials}\ }\textbf {\bibinfo {volume} {9}},\
  \bibinfo {pages} {1901597} (\bibinfo {year} {2019})}\BibitemShut {NoStop}%
\bibitem [{\citenamefont {Edstr{\"o}m}\ \emph {et~al.}(2006)\citenamefont
  {Edstr{\"o}m}, \citenamefont {Herstedt},\ and\ \citenamefont
  {Abraham}}]{edstrom2006new}%
  \BibitemOpen
  \bibfield  {author} {\bibinfo {author} {\bibfnamefont {K.}~\bibnamefont
  {Edstr{\"o}m}}, \bibinfo {author} {\bibfnamefont {M.}~\bibnamefont
  {Herstedt}},\ and\ \bibinfo {author} {\bibfnamefont {D.~P.}\ \bibnamefont
  {Abraham}},\ }\bibfield  {title} {\bibinfo {title} {A new look at the solid
  electrolyte interphase on graphite anodes in li-ion batteries},\ }\href@noop
  {} {\bibfield  {journal} {\bibinfo  {journal} {Journal of Power Sources}\
  }\textbf {\bibinfo {volume} {153}},\ \bibinfo {pages} {380} (\bibinfo {year}
  {2006})}\BibitemShut {NoStop}%
\bibitem [{\citenamefont {Huang}\ \emph {et~al.}(2019)\citenamefont {Huang},
  \citenamefont {Attia}, \citenamefont {Wang}, \citenamefont {Renfrew},
  \citenamefont {Jin}, \citenamefont {Das}, \citenamefont {Zhang},
  \citenamefont {Boyle}, \citenamefont {Li}, \citenamefont {Bazant} \emph
  {et~al.}}]{huang2019evolution}%
  \BibitemOpen
  \bibfield  {author} {\bibinfo {author} {\bibfnamefont {W.}~\bibnamefont
  {Huang}}, \bibinfo {author} {\bibfnamefont {P.~M.}\ \bibnamefont {Attia}},
  \bibinfo {author} {\bibfnamefont {H.}~\bibnamefont {Wang}}, \bibinfo {author}
  {\bibfnamefont {S.~E.}\ \bibnamefont {Renfrew}}, \bibinfo {author}
  {\bibfnamefont {N.}~\bibnamefont {Jin}}, \bibinfo {author} {\bibfnamefont
  {S.}~\bibnamefont {Das}}, \bibinfo {author} {\bibfnamefont {Z.}~\bibnamefont
  {Zhang}}, \bibinfo {author} {\bibfnamefont {D.~T.}\ \bibnamefont {Boyle}},
  \bibinfo {author} {\bibfnamefont {Y.}~\bibnamefont {Li}}, \bibinfo {author}
  {\bibfnamefont {M.~Z.}\ \bibnamefont {Bazant}}, \emph {et~al.},\ }\bibfield
  {title} {\bibinfo {title} {Evolution of the solid--electrolyte interphase on
  carbonaceous anodes visualized by atomic-resolution cryogenic electron
  microscopy},\ }\href@noop {} {\bibfield  {journal} {\bibinfo  {journal} {Nano
  letters}\ }\textbf {\bibinfo {volume} {19}},\ \bibinfo {pages} {5140}
  (\bibinfo {year} {2019})}\BibitemShut {NoStop}%
\bibitem [{\citenamefont {Omar}\ and\ \citenamefont
  {Ahmad}(2016)}]{omar2016electrical}%
  \BibitemOpen
  \bibfield  {author} {\bibinfo {author} {\bibfnamefont {M.~K.}\ \bibnamefont
  {Omar}}\ and\ \bibinfo {author} {\bibfnamefont {A.~H.}\ \bibnamefont
  {Ahmad}},\ }\bibfield  {title} {\bibinfo {title} {Electrical impedance
  spectroscopy and fourier transform infrared studies of new binary li2co3-lii
  solid electrolyte},\ }in\ \href@noop {} {\emph {\bibinfo {booktitle}
  {Materials Science Forum}}},\ Vol.\ \bibinfo {volume} {846}\ (\bibinfo
  {organization} {Trans Tech Publ},\ \bibinfo {year} {2016})\ pp.\ \bibinfo
  {pages} {517--522}\BibitemShut {NoStop}%
\bibitem [{\citenamefont {Zhuang}\ and\ \citenamefont
  {Bazant}(2022)}]{zhuang2022theory}%
  \BibitemOpen
  \bibfield  {author} {\bibinfo {author} {\bibfnamefont {D.}~\bibnamefont
  {Zhuang}}\ and\ \bibinfo {author} {\bibfnamefont {M.~Z.}\ \bibnamefont
  {Bazant}},\ }\bibfield  {title} {\bibinfo {title} {Theory of layered-oxide
  cathode degradation in li-ion batteries by oxidation-induced cation
  disorder},\ }\href@noop {} {\bibfield  {journal} {\bibinfo  {journal}
  {Journal of The Electrochemical Society}\ } (\bibinfo {year}
  {2022})}\BibitemShut {NoStop}%
\bibitem [{\citenamefont {Lin}\ \emph {et~al.}(2014)\citenamefont {Lin},
  \citenamefont {Markus}, \citenamefont {Nordlund}, \citenamefont {Weng},
  \citenamefont {Asta}, \citenamefont {Xin},\ and\ \citenamefont
  {Doeff}}]{lin2014surface}%
  \BibitemOpen
  \bibfield  {author} {\bibinfo {author} {\bibfnamefont {F.}~\bibnamefont
  {Lin}}, \bibinfo {author} {\bibfnamefont {I.~M.}\ \bibnamefont {Markus}},
  \bibinfo {author} {\bibfnamefont {D.}~\bibnamefont {Nordlund}}, \bibinfo
  {author} {\bibfnamefont {T.-C.}\ \bibnamefont {Weng}}, \bibinfo {author}
  {\bibfnamefont {M.~D.}\ \bibnamefont {Asta}}, \bibinfo {author}
  {\bibfnamefont {H.~L.}\ \bibnamefont {Xin}},\ and\ \bibinfo {author}
  {\bibfnamefont {M.~M.}\ \bibnamefont {Doeff}},\ }\bibfield  {title} {\bibinfo
  {title} {Surface reconstruction and chemical evolution of stoichiometric
  layered cathode materials for lithium-ion batteries},\ }\href@noop {}
  {\bibfield  {journal} {\bibinfo  {journal} {Nature communications}\ }\textbf
  {\bibinfo {volume} {5}},\ \bibinfo {pages} {1} (\bibinfo {year}
  {2014})}\BibitemShut {NoStop}%
\bibitem [{\citenamefont {Yan}\ \emph {et~al.}(2017)\citenamefont {Yan},
  \citenamefont {Zheng}, \citenamefont {Zhang},\ and\ \citenamefont
  {Wang}}]{yan2017atomic}%
  \BibitemOpen
  \bibfield  {author} {\bibinfo {author} {\bibfnamefont {P.}~\bibnamefont
  {Yan}}, \bibinfo {author} {\bibfnamefont {J.}~\bibnamefont {Zheng}}, \bibinfo
  {author} {\bibfnamefont {J.-G.}\ \bibnamefont {Zhang}},\ and\ \bibinfo
  {author} {\bibfnamefont {C.}~\bibnamefont {Wang}},\ }\bibfield  {title}
  {\bibinfo {title} {Atomic resolution structural and chemical imaging
  revealing the sequential migration of ni, co, and mn upon the battery cycling
  of layered cathode},\ }\href@noop {} {\bibfield  {journal} {\bibinfo
  {journal} {Nano letters}\ }\textbf {\bibinfo {volume} {17}},\ \bibinfo
  {pages} {3946} (\bibinfo {year} {2017})}\BibitemShut {NoStop}%
\bibitem [{\citenamefont {Zheng}\ \emph {et~al.}(2015)\citenamefont {Zheng},
  \citenamefont {Tan}, \citenamefont {Zhang}, \citenamefont {Qu}, \citenamefont
  {Wan}, \citenamefont {Wang}, \citenamefont {Shen},\ and\ \citenamefont
  {Zheng}}]{zheng2015correlation}%
  \BibitemOpen
  \bibfield  {author} {\bibinfo {author} {\bibfnamefont {H.}~\bibnamefont
  {Zheng}}, \bibinfo {author} {\bibfnamefont {L.}~\bibnamefont {Tan}}, \bibinfo
  {author} {\bibfnamefont {L.}~\bibnamefont {Zhang}}, \bibinfo {author}
  {\bibfnamefont {Q.}~\bibnamefont {Qu}}, \bibinfo {author} {\bibfnamefont
  {Z.}~\bibnamefont {Wan}}, \bibinfo {author} {\bibfnamefont {Y.}~\bibnamefont
  {Wang}}, \bibinfo {author} {\bibfnamefont {M.}~\bibnamefont {Shen}},\ and\
  \bibinfo {author} {\bibfnamefont {H.}~\bibnamefont {Zheng}},\ }\bibfield
  {title} {\bibinfo {title} {Correlation between lithium deposition on graphite
  electrode and the capacity loss for lifepo4/graphite cells},\ }\href@noop {}
  {\bibfield  {journal} {\bibinfo  {journal} {Electrochimica Acta}\ }\textbf
  {\bibinfo {volume} {173}},\ \bibinfo {pages} {323} (\bibinfo {year}
  {2015})}\BibitemShut {NoStop}%
\bibitem [{\citenamefont {An}\ \emph {et~al.}(2016)\citenamefont {An},
  \citenamefont {Li}, \citenamefont {Daniel}, \citenamefont {Mohanty},
  \citenamefont {Nagpure},\ and\ \citenamefont {Wood~III}}]{an2016state}%
  \BibitemOpen
  \bibfield  {author} {\bibinfo {author} {\bibfnamefont {S.~J.}\ \bibnamefont
  {An}}, \bibinfo {author} {\bibfnamefont {J.}~\bibnamefont {Li}}, \bibinfo
  {author} {\bibfnamefont {C.}~\bibnamefont {Daniel}}, \bibinfo {author}
  {\bibfnamefont {D.}~\bibnamefont {Mohanty}}, \bibinfo {author} {\bibfnamefont
  {S.}~\bibnamefont {Nagpure}},\ and\ \bibinfo {author} {\bibfnamefont {D.~L.}\
  \bibnamefont {Wood~III}},\ }\bibfield  {title} {\bibinfo {title} {The state
  of understanding of the lithium-ion-battery graphite solid electrolyte
  interphase (sei) and its relationship to formation cycling},\ }\href@noop {}
  {\bibfield  {journal} {\bibinfo  {journal} {Carbon}\ }\textbf {\bibinfo
  {volume} {105}},\ \bibinfo {pages} {52} (\bibinfo {year} {2016})}\BibitemShut
  {NoStop}%
\bibitem [{\citenamefont {An}\ \emph {et~al.}(2017)\citenamefont {An},
  \citenamefont {Li}, \citenamefont {Du}, \citenamefont {Daniel},\ and\
  \citenamefont {Wood~III}}]{an2017fast}%
  \BibitemOpen
  \bibfield  {author} {\bibinfo {author} {\bibfnamefont {S.~J.}\ \bibnamefont
  {An}}, \bibinfo {author} {\bibfnamefont {J.}~\bibnamefont {Li}}, \bibinfo
  {author} {\bibfnamefont {Z.}~\bibnamefont {Du}}, \bibinfo {author}
  {\bibfnamefont {C.}~\bibnamefont {Daniel}},\ and\ \bibinfo {author}
  {\bibfnamefont {D.~L.}\ \bibnamefont {Wood~III}},\ }\bibfield  {title}
  {\bibinfo {title} {Fast formation cycling for lithium ion batteries},\
  }\href@noop {} {\bibfield  {journal} {\bibinfo  {journal} {Journal of Power
  Sources}\ }\textbf {\bibinfo {volume} {342}},\ \bibinfo {pages} {846}
  (\bibinfo {year} {2017})}\BibitemShut {NoStop}%
\bibitem [{\citenamefont {Sacci}\ \emph {et~al.}(2014)\citenamefont {Sacci},
  \citenamefont {Dudney}, \citenamefont {More}, \citenamefont {Parent},
  \citenamefont {Arslan}, \citenamefont {Browning},\ and\ \citenamefont
  {Unocic}}]{sacci2014direct}%
  \BibitemOpen
  \bibfield  {author} {\bibinfo {author} {\bibfnamefont {R.~L.}\ \bibnamefont
  {Sacci}}, \bibinfo {author} {\bibfnamefont {N.~J.}\ \bibnamefont {Dudney}},
  \bibinfo {author} {\bibfnamefont {K.~L.}\ \bibnamefont {More}}, \bibinfo
  {author} {\bibfnamefont {L.~R.}\ \bibnamefont {Parent}}, \bibinfo {author}
  {\bibfnamefont {I.}~\bibnamefont {Arslan}}, \bibinfo {author} {\bibfnamefont
  {N.~D.}\ \bibnamefont {Browning}},\ and\ \bibinfo {author} {\bibfnamefont
  {R.~R.}\ \bibnamefont {Unocic}},\ }\bibfield  {title} {\bibinfo {title}
  {Direct visualization of initial sei morphology and growth kinetics during
  lithium deposition by in situ electrochemical transmission electron
  microscopy},\ }\href@noop {} {\bibfield  {journal} {\bibinfo  {journal}
  {Chemical Communications}\ }\textbf {\bibinfo {volume} {50}},\ \bibinfo
  {pages} {2104} (\bibinfo {year} {2014})}\BibitemShut {NoStop}%
\bibitem [{\citenamefont {Aykol}\ \emph {et~al.}(2021)\citenamefont {Aykol},
  \citenamefont {Gopal}, \citenamefont {Anapolsky}, \citenamefont {Herring},
  \citenamefont {van Vlijmen}, \citenamefont {Berliner}, \citenamefont
  {Bazant}, \citenamefont {Braatz}, \citenamefont {Chueh},\ and\ \citenamefont
  {Storey}}]{aykol2021perspective}%
  \BibitemOpen
  \bibfield  {author} {\bibinfo {author} {\bibfnamefont {M.}~\bibnamefont
  {Aykol}}, \bibinfo {author} {\bibfnamefont {C.~B.}\ \bibnamefont {Gopal}},
  \bibinfo {author} {\bibfnamefont {A.}~\bibnamefont {Anapolsky}}, \bibinfo
  {author} {\bibfnamefont {P.~K.}\ \bibnamefont {Herring}}, \bibinfo {author}
  {\bibfnamefont {B.}~\bibnamefont {van Vlijmen}}, \bibinfo {author}
  {\bibfnamefont {M.~D.}\ \bibnamefont {Berliner}}, \bibinfo {author}
  {\bibfnamefont {M.~Z.}\ \bibnamefont {Bazant}}, \bibinfo {author}
  {\bibfnamefont {R.~D.}\ \bibnamefont {Braatz}}, \bibinfo {author}
  {\bibfnamefont {W.~C.}\ \bibnamefont {Chueh}},\ and\ \bibinfo {author}
  {\bibfnamefont {B.~D.}\ \bibnamefont {Storey}},\ }\bibfield  {title}
  {\bibinfo {title} {Perspective—combining physics and machine learning to
  predict battery lifetime},\ }\href@noop {} {\bibfield  {journal} {\bibinfo
  {journal} {Journal of The Electrochemical Society}\ }\textbf {\bibinfo
  {volume} {168}},\ \bibinfo {pages} {030525} (\bibinfo {year}
  {2021})}\BibitemShut {NoStop}%
\bibitem [{\citenamefont {Zwanzig}(2001)}]{zwanzig2001nonequilibrium}%
  \BibitemOpen
  \bibfield  {author} {\bibinfo {author} {\bibfnamefont {R.}~\bibnamefont
  {Zwanzig}},\ }\href@noop {} {\emph {\bibinfo {title} {Nonequilibrium
  statistical mechanics}}}\ (\bibinfo  {publisher} {Oxford university press},\
  \bibinfo {year} {2001})\BibitemShut {NoStop}%
\bibitem [{\citenamefont {Kondepudi}\ and\ \citenamefont
  {Prigogine}(2014)}]{kondepudi2014modern}%
  \BibitemOpen
  \bibfield  {author} {\bibinfo {author} {\bibfnamefont {D.}~\bibnamefont
  {Kondepudi}}\ and\ \bibinfo {author} {\bibfnamefont {I.}~\bibnamefont
  {Prigogine}},\ }\href@noop {} {\emph {\bibinfo {title} {Modern
  thermodynamics: from heat engines to dissipative structures}}}\ (\bibinfo
  {publisher} {John Wiley \& Sons},\ \bibinfo {year} {2014})\BibitemShut
  {NoStop}%
\bibitem [{\citenamefont {Li}\ \emph {et~al.}(2018{\natexlab{b}})\citenamefont
  {Li}, \citenamefont {Chen}, \citenamefont {Lim}, \citenamefont {Deng},
  \citenamefont {Lim}, \citenamefont {Fraggedakis}, \citenamefont {Attia},
  \citenamefont {Lee}, \citenamefont {Jin}, \citenamefont {Mo{\v{s}}kon} \emph
  {et~al.}}]{li2018fluid}%
  \BibitemOpen
  \bibfield  {author} {\bibinfo {author} {\bibfnamefont {Y.}~\bibnamefont
  {Li}}, \bibinfo {author} {\bibfnamefont {H.}~\bibnamefont {Chen}}, \bibinfo
  {author} {\bibfnamefont {K.}~\bibnamefont {Lim}}, \bibinfo {author}
  {\bibfnamefont {H.~D.}\ \bibnamefont {Deng}}, \bibinfo {author}
  {\bibfnamefont {J.}~\bibnamefont {Lim}}, \bibinfo {author} {\bibfnamefont
  {D.}~\bibnamefont {Fraggedakis}}, \bibinfo {author} {\bibfnamefont {P.~M.}\
  \bibnamefont {Attia}}, \bibinfo {author} {\bibfnamefont {S.~C.}\ \bibnamefont
  {Lee}}, \bibinfo {author} {\bibfnamefont {N.}~\bibnamefont {Jin}}, \bibinfo
  {author} {\bibfnamefont {J.}~\bibnamefont {Mo{\v{s}}kon}}, \emph {et~al.},\
  }\bibfield  {title} {\bibinfo {title} {Fluid-enhanced surface diffusion
  controls intraparticle phase transformations},\ }\href@noop {} {\bibfield
  {journal} {\bibinfo  {journal} {Nature materials}\ }\textbf {\bibinfo
  {volume} {17}},\ \bibinfo {pages} {915} (\bibinfo {year}
  {2018}{\natexlab{b}})}\BibitemShut {NoStop}%
\bibitem [{\citenamefont {Risken}(1996)}]{risken1996fokker}%
  \BibitemOpen
  \bibfield  {author} {\bibinfo {author} {\bibfnamefont {H.}~\bibnamefont
  {Risken}},\ }\bibfield  {title} {\bibinfo {title} {Fokker-planck equation},\
  }in\ \href@noop {} {\emph {\bibinfo {booktitle} {The Fokker-Planck
  Equation}}}\ (\bibinfo  {publisher} {Springer},\ \bibinfo {year} {1996})\
  pp.\ \bibinfo {pages} {63--95}\BibitemShut {NoStop}%
\bibitem [{\citenamefont {Colclasure}\ \emph {et~al.}(2020)\citenamefont
  {Colclasure}, \citenamefont {Tanim}, \citenamefont {Jansen}, \citenamefont
  {Trask}, \citenamefont {Dunlop}, \citenamefont {Polzin}, \citenamefont
  {Bloom}, \citenamefont {Robertson}, \citenamefont {Flores}, \citenamefont
  {Evans} \emph {et~al.}}]{colclasure2020electrode}%
  \BibitemOpen
  \bibfield  {author} {\bibinfo {author} {\bibfnamefont {A.~M.}\ \bibnamefont
  {Colclasure}}, \bibinfo {author} {\bibfnamefont {T.~R.}\ \bibnamefont
  {Tanim}}, \bibinfo {author} {\bibfnamefont {A.~N.}\ \bibnamefont {Jansen}},
  \bibinfo {author} {\bibfnamefont {S.~E.}\ \bibnamefont {Trask}}, \bibinfo
  {author} {\bibfnamefont {A.~R.}\ \bibnamefont {Dunlop}}, \bibinfo {author}
  {\bibfnamefont {B.~J.}\ \bibnamefont {Polzin}}, \bibinfo {author}
  {\bibfnamefont {I.}~\bibnamefont {Bloom}}, \bibinfo {author} {\bibfnamefont
  {D.}~\bibnamefont {Robertson}}, \bibinfo {author} {\bibfnamefont
  {L.}~\bibnamefont {Flores}}, \bibinfo {author} {\bibfnamefont
  {M.}~\bibnamefont {Evans}}, \emph {et~al.},\ }\bibfield  {title} {\bibinfo
  {title} {Electrode scale and electrolyte transport effects on extreme fast
  charging of lithium-ion cells},\ }\href@noop {} {\bibfield  {journal}
  {\bibinfo  {journal} {Electrochimica Acta}\ }\textbf {\bibinfo {volume}
  {337}},\ \bibinfo {pages} {135854} (\bibinfo {year} {2020})}\BibitemShut
  {NoStop}%
\end{thebibliography}%

\end{document}